\documentclass[reprint,superscriptaddress,amsmath,amssymb,aps,longbibliography]{revtex4-2}
\usepackage{graphicx}
\usepackage[version=4]{mhchem}
\usepackage{booktabs}
\usepackage{pdfpages}
\usepackage[hidelinks]{hyperref}
\hypersetup{pdftitle={Using large language models to probe the limits of atom-centered structural descriptors},pdfauthor={Michelangelo Domina and Michele Ceriotti}}
\newcommand{\br}{\boldsymbol{r}}
\IfFileExists{arxiv/text.tex}{\def\arxivpath{arxiv/}}{\def\arxivpath{}}
\makeatletter
\AtBeginDocument{\let\LS@rot\@undefined}
\makeatother

\begin{document}
\title{\texorpdfstring{Using large language models to probe the limits of\\ atom-centered structural descriptors}{Using large language models to probe the limits of atom-centered structural descriptors}}
\author{Michelangelo Domina}
\affiliation{Laboratory of Computational Science and Modeling, Institut des Mat\'eriaux, \'Ecole Polytechnique F\'ed\'erale de Lausanne, 1015 Lausanne, Switzerland}
\author{Michele Ceriotti}
\email{michele.ceriotti@epfl.ch}
\affiliation{Laboratory of Computational Science and Modeling, Institut des Mat\'eriaux, \'Ecole Polytechnique F\'ed\'erale de Lausanne, 1015 Lausanne, Switzerland}
\date{October 2, 2026}

\begin{abstract}
Mapping an atomic structure to a compact set of geometric descriptors is an essential step in any machine-learning application to atomic-scale modeling.
A powerful and widely used approach can be understood as a discretization of the histogram of pair distances, triangles, etc., which results in a hierarchy of symmetry-invariant atom-centered descriptors.
Unfortunately, the lower rungs on this hierarchy (two-, three-, and four-body clusters) were found to be incomplete, with symmetry-unrelated pairs of structures having exactly the same descriptors.
However, all the ``degeneracies'' reported so far are resolved by considering larger clusters of neighbors to build the descriptors.
We report examples of 3D structures that are indistinguishable even if one considers clusters of up to \emph{seven} neighbors, and to arbitrary order when considering a practical level of discretization of the descriptors, discovered with the assistance of large language models. The key ingredients in their construction can be traced to results that have been known for decades in different communities; the model was able to find the references and recognize their significance for the problem at hand.
We believe that this experiment exposes an extremely fruitful usage pattern for Artificial Intelligence in science: translating results between different communities and application domains, accelerating the process by which serendipitous discoveries in a field become breakthroughs in another.
\end{abstract}
\maketitle

\section{Introduction}

Machine-learning models used to classify, or predict properties of, molecules and condensed-matter groups of atoms can be seen as operating over 3D point clouds, decorated with labels describing the chemical nature of the atoms.
A central focus of the early days of this field (which is worth remembering dates back at least to the early 2000s~\cite{behl-parr07prl,rupp+12prl,bart+10prl}) involved building local, often atom-centered, \emph{representations} of configurations that encode geometric relations between atoms in a symmetry-adapted form~\cite{musi+21cr}; these local representations were then used as inputs to machine-learning models for interatomic potentials~\cite{behl21cr,unke+21cr}, or for electronic-structure information such as valence electron densities and effective tight-binding Hamiltonians~\cite{gris+19acscs,deeptb}.
Even though many alternative approaches were proposed to build these representations, a systematic study revealed that constraints imposed by the symmetry structure of the $O(3)$ group restrict the operations one can use to construct symmetric representations, and so all approaches could be systematically classified in terms of a hierarchy of $\nu$-point correlations of the atomic neighbor density~\cite{will+19jcp}.
Almost at the same time, a few papers published by Pozdnyakov et al.~\cite{pozd+20prl,pozd+21ore,pozd+22mlst} revealed, to the surprise of many, that the simplest, widely used 2-, 3-, and 4-body correlations are \emph{incomplete}, i.e. that there are pairs of atomic arrangements that are not related by symmetry, but share the exact same correlations, irrespective of the radial cutoff, or the resolution used to discretize the neighbor density.

The practical implications of these degeneracies are not conclusively established: for a fixed number of neighbors, exact three-body degeneracies lie on lower-dimensional sets~\cite{pozd+20prl}, even though numerical instabilities affect configurations that are near-degenerate~\cite{pozd+21ore,pars-goed22jcp,pozd+22jcp}, and the most direct evidence of their impact in a realistic application only becomes apparent in the regime of a very large dataset of very high-energy configurations. The distinction between generic and universal completeness also appears in a related global problem: according to Boutin and Kemper's dimension and size conditions, the pair-distance multiset determines almost every finite point cloud up to congruence, but exact counterexamples exist~\cite{bout-kemp04aam}. Nevertheless, the realization that low-order representations have fundamental limitations played a role in accelerating the transition to using higher-order correlations~\cite{drau19prb,niga+20jcp}, to a concerted effort to define provably complete descriptors~\cite{duss+22jcp,niga+24aplml,maen+24jpcl} as well as to the use of equivariant neural networks that build learnable high-order correlations, and include equivariant message passing steps that further increase the resolving power of the models~\cite{bata+22nips,bazt+22ncomm}.
Several questions remained open: is the trispectrum (5-body correlations) sufficient to resolve all pairs of atomic environments? Are there structures that are degenerate relative to all their atom-centered environments, rather than just one, as is the case for distance-based graph networks~\cite{pozd+22mlst}? Even though these might be little more than curiosities for a field that is starting to question whether rotational equivariance must be hard-wired into the backbone architecture~\cite{pozd-ceri23nips,NEURIPS2024_fad8e191}, some of us kept thinking about them -- admittedly without much sense of urgency -- but could not make any progress beyond the known examples from Pozdnyakov et al.

In recent months, several reports have described LLM-assisted progress on difficult problems in pure and applied mathematics~\cite{balko2026bolzanocasestudiesllmassisted,feng2026semiautonomousmathematicsdiscoverygemini}, including several listed as open in the Erd\H{o}s Problems database~\cite{feng2026semiautonomousmathematicsdiscoverygemini}. Sawin subsequently gave an explicit quantitative refinement of a separate LLM-assisted disproof~\cite{sawin2026explicitlowerboundunit}.
This got us curious as to whether they would be similarly helpful in attacking this problem, which, while being clearly mathematical in nature, has some relevance to the use of geometric machine learning in chemical physics and materials simulations.
We report on our experiments in this direction, not only because the outcomes are positive (the LLMs proposed several geometric examples that are degenerate, not only for the trispectrum but up to the hexaspectrum, and up to arbitrary order when using a finite angular discretization of the neighbor density), but because we believe that the process (\emph{how} the LLM achieved these results) provides useful lessons as to how this community could leverage existing, publicly accessible, general-purpose LLMs as assistants to accelerate scientific discovery.

\section{Atom-centered descriptors}

Atom-centered descriptors represent an atomic configuration as a collection of local environments, expressed in a form that is invariant to translations, rotations, and permutations of equivalent atoms~\cite{bart+13prb,will+19jcp,drau19prb}. For the environment around atom $i$, we define the neighbor density from the relative positions $\br_{ij}=\br_j-\br_i$ as
\begin{equation}
    \rho_i(\br) = \sum_j w_j f_\text{cut}(r_{ij}) \delta(\br-\br_{ij}),
\end{equation}
where $r_{ij}=|\br_{ij}|$ -- which is invariant to translations and neighbor-index permutations.
The cutoff function enforces locality, while the weights $w_j$ distinguish atomic species or, more abstractly, color the points in the cloud. The weights are usually vectors, and can be implemented in practice as a one-hot encoding of elements or an alchemical contraction~\cite{will+18pccp}. In what follows we assume, for simplicity and because multiple density channels usually \emph{lift} geometric degeneracies, $w_j=1$.

In most implementations the point density is usually smoothed, and/or discretized on a finite radial and angular basis
\begin{equation}
    c^{(i)}_{nl m}
    =
    \int \mathrm d\br\,
    R_n(r)Y_{l m}^*(\hat{\br})\rho_i(\br),
\end{equation}
where $R_n$ are radial basis functions and $Y_{l m}$ are spherical harmonics. Rotational invariants are then constructed by coupling products of these coefficients to total angular momentum zero, and to even parity when invariance under the full $O(3)$ group is required.

The density correlations are usually evaluated directly in this basis. Products of $\nu$ expansion coefficients are coupled with Clebsch--Gordan coefficients so that the resulting feature transforms as a scalar,
\begin{equation}
    B^{(i)}_{\nu\boldsymbol n\boldsymbol l}
    =
    \sum_{m_1\ldots m_\nu}
    C^{0}_{l_1 m_1 \ldots l_\nu m_\nu}
    \prod_{a=1}^{\nu} c^{(i)}_{n_al_a m_a},
\end{equation}
where $C^{0}_{l_1 m_1 \ldots l_\nu m_\nu}$ denotes a generalized angular-momentum coupling to total angular momentum zero, and $\boldsymbol n$ and $\boldsymbol l$ collect the indices of the radial and angular basis functions, respectively. These scalar contractions are the rotationally invariant $\nu$-correlations of the neighbor density. In signal-processing language, $\nu=2$ corresponds to the power spectrum (introduced in atomistic modeling as the Smooth Overlap of Atomic Positions, SOAP~\cite{bart+13prb}), $\nu=3$ to the bispectrum (used in the spectral neighbor analysis potential, SNAP~\cite{thom+15jcp}), $\nu=4$ to the trispectrum, and so on.
The full, untruncated hierarchy provides a complete basis for expanding symmetry-invariant functions of the neighbor coordinates~\cite{domina+prb25,duss+22jcp}. At fixed order $\nu$, the correlations encode the geometries of $\nu$ neighbors through their distances from the center and their mutual angles. Counting the central atom, they correspond to $(\nu+1)$-body correlations: $\nu=1$ features characterize the set of pair distances, $\nu=2$ the set of triangles, and $\nu=3$ the set of tetrahedra. At this order, parity-odd $SO(3)$-invariant bispectrum components can distinguish a chiral tetrahedron from its mirror image, whereas an $O(3)$-invariant representation omits these components because they change sign under reflection~\cite{kaka12jmiv}.
Higher-order invariants are obtained by coupling products of four and more coefficients, making the descriptor sensitive to progressively larger local arrangements.

\begin{figure}
    \centering
\includegraphics[width=\linewidth]{\arxivpath 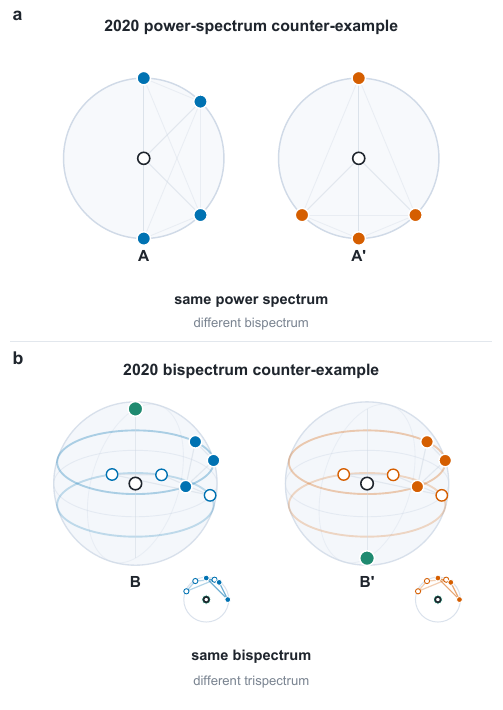}
    \caption{Atomic configurations that (a) share the same power spectrum (SOAP features, 3-body symmetry functions) and (b) share the same bispectrum (SNAP descriptors), as presented in Ref.~\citenum{pozd+20prl}.}
    \label{fig:pozdnyakov}
\end{figure}

Because $\nu=1$ retains only the multiset of center--neighbor radii, it cannot uniquely characterize an environment. A classical global analogue arises in crystallography: the Patterson function~\cite{patt39nature} is the autocorrelation of a full structure density, equivalently the weighted distribution of all interatomic displacements. It pools one-neighbor information over all atoms taken in turn as the center and therefore corresponds to $\nu=1$ in the present center-conditioned hierarchy, even though it is quadratic in the global density. Patterson introduced the term \emph{homometric} for structures with the same species-resolved Patterson function and hence identical diffraction patterns, emphasizing that they need be neither identical nor mirror-related. Rosenblatt and Seymour later gave a complete algebraic characterization of Patterson homometry for finite point distributions with integer weights~\cite{rose-seym82siam}. By contrast, $\nu=2$ correlates two neighbors about a retained center and encodes triangles. The successful use of the SOAP power spectrum~\cite{bart+13prb} and related three-body atom-centered symmetry functions~\cite{behl11jcp} suggested much stronger resolving power.
However, in the early 2020s Pozdnyakov et al. proposed a geometric construction~\cite{pozd+20prl} yielding \emph{degenerate} pairs that cannot be distinguished by the power spectrum, and one for structures that cannot be distinguished by the bispectrum~(see Fig.~\ref{fig:pozdnyakov}).

The existence of these pathological configurations implies a limitation of any model built on them, even though the extent to which it affects performance in practical cases is difficult to quantify.
Ref.~\citenum{pozd+20prl} demonstrates a clear degradation of model performance for a potential for \ce{CH4} configurations based on the power spectrum (even if the dataset does not contain specifically power-spectrum-breaking pairs), but in more complicated cases it is difficult to tell whether poor performance of low-order invariant models depends specifically on incompleteness rather than on other architectural factors.
In addition, there are many ways an ML model can be modified to address these limiting cases. First, any global property can be expressed as a sum of contributions from each atom, and so the \emph{neighbor}-centered environments can be used to lift the degeneracies.
Second, neighbor-centered information can be used in a message-passing scheme to also resolve the ambiguity at the environment level (pair-distance-limited features being the only case for which a true structure-level degeneracy has been found~\cite{pozd+22mlst}).
Third, increasing the descriptor order can resolve individual degeneracies: the bispectrum distinguishes the power-spectrum pair in Fig.~\ref{fig:pozdnyakov}(a), and the trispectrum distinguishes the bispectrum pair in Fig.~\ref{fig:pozdnyakov}(b)~\cite{pozd+20prl,validation_archive}.

It is also worth mentioning that alternative frameworks exist that achieve completeness with low-order invariants following alternative approaches.
These include, e.g., defining a common reference frame by aligning the tensors of inertia~\cite{kurlin2024polynomial}, using the centroid and pairs of selected points as common references~\cite{kurlin2026complete}, or defining scalar descriptors relative to triplets of atoms rather than individual centers~\cite{niga+24aplml}.

In all these and in similar constructions~\cite{widd-kurl22nips,hord+24aaai}, features remain tied to a shared center, reference frame, or ordered tuple, whereas fixed-order density correlations pool small motifs without recording which motifs share points. The question pursued below is therefore more specific: how far can incompleteness persist within the familiar hierarchy of fixed-order density correlations?

\begin{figure*}
    \centering
\includegraphics[width=\linewidth]{\arxivpath 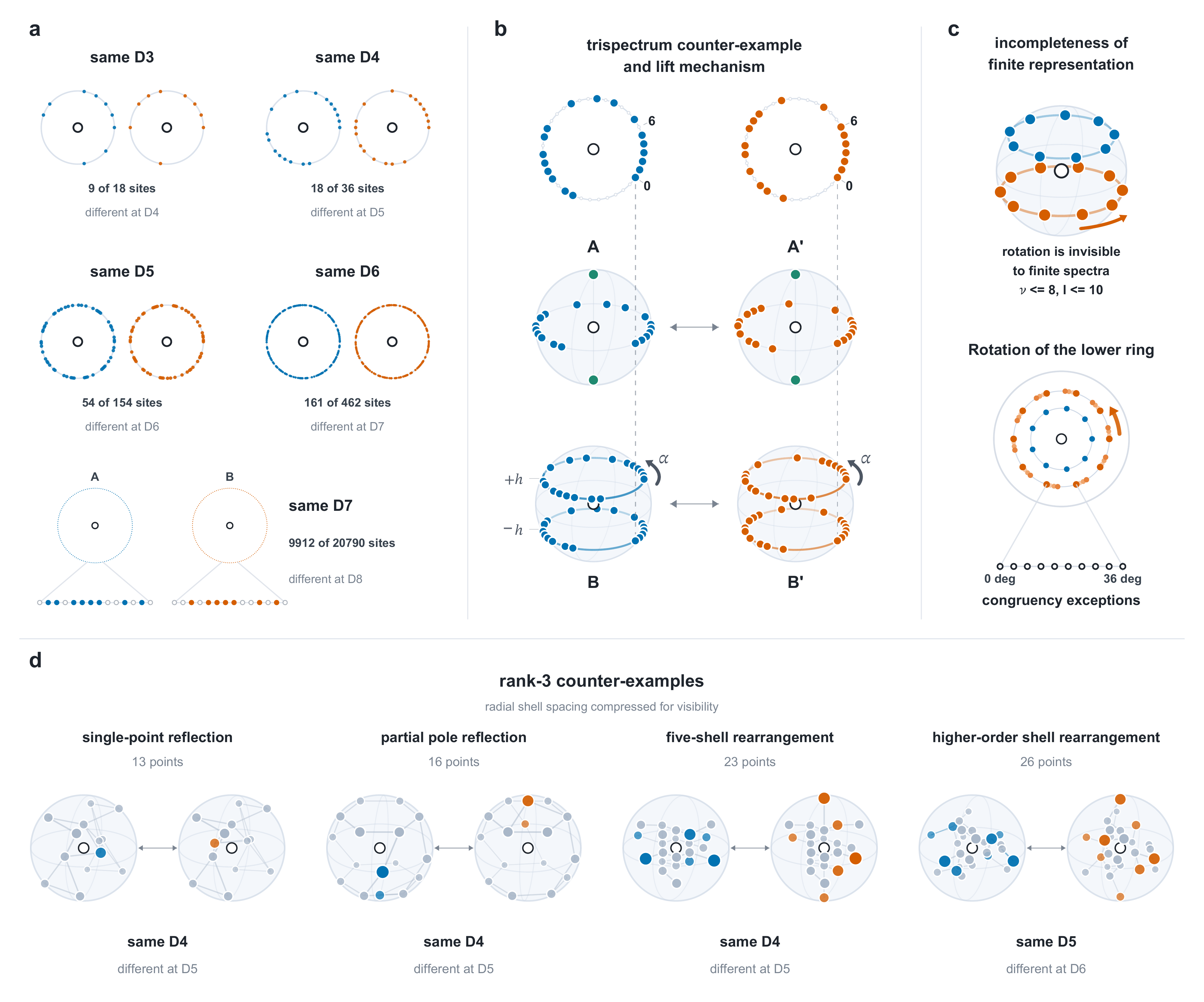}
    \caption{Atlas of representation-degenerate atom-centered environments. The blue and orange circles distinguish the two members of each pair, gray marks points common to both, and the open circle is the central atom. (a) Planar cyclic pairs with identical complete decks through the order shown. (b) Two three-dimensional lifts of the 18-of-36 cyclic pair: adding two polar atoms gives $A,A'$, while copying the pattern onto two parallel rings gives $B,B'$. Rotating the upper ring by the same arbitrary angle $\alpha$ in both environments preserves the $D_4$ degeneracy. (c) Finite-representation degeneracy for coaxial 9- and 10-fold rings. Every relative angle has the same retained features for $\nu_{\max}=8$ and $l_{\max}=10$. The markers identify the exceptional angles at which the point clouds are $O(3)$ congruent. (d) Intrinsic rank-three pairs whose degeneracy is genuinely three-dimensional. For reflection-even $O(3)$ invariants, the first three pairs agree through $D_4$ and the last through $D_5$. Parity-odd $SO(3)$ invariants distinguish the 13- and 16-point pairs at $D_4$, and the 23- and 26-point pairs at $D_3$. The radial-shell separations are compressed for visibility.}
    \label{fig:counterexample-atlas}
\end{figure*}

\section{Degenerate structures for high-order correlations}

Even though $\nu=4$ descriptors suffice to lift the degeneracy of the structures in Fig.~\ref{fig:pozdnyakov},  there are pairs of atomic configurations that cannot be resolved by the trispectrum -- nor by the $\nu=5,6,7$ correlations.
In fact, if one also considers a finite discretization level of the angular correlations (which is always the case in practice) one can build counterexamples for \emph{any correlation order}.
We found these configurations, which we will discuss in this section, with the help of a large language model. We want, however, to first provide an atlas of the different types of degeneracies that were identified, sketching the mechanism that determines loss of resolving power, and leaving detailed constructions, coordinates, validation procedures, and search prompts to the supplementary material.
\nocite{kurlin2025periodic,simo17phd,pebo07cpc,deng-mood09jsp,deng-mood12cmb}

\subsection{\texorpdfstring{$\nu$}{nu}-decks and cyclic structures}
\label{sec:cyclic-decks}

Take a regular polygon with $N$ sites and index its vertices by $s=0,\ldots,N-1$, with addition understood modulo $N$.
A subset $S\subset\mathbb Z_N$ specifies which sites are occupied.
To turn this pattern into an atom-centered environment, place the central atom at the origin and, for every $s\in S$, a neighbor at azimuth $\phi_s=2\pi s/N$, i.e. at $\br_s=R(\cos\phi_s,\sin\phi_s,0)$.
All neighbors then lie on the ring of radius $R$ shown in Fig.~\ref{fig:counterexample-atlas}(a); different environments correspond to different choices of $S$.
The integer $s$ is therefore just a site label, and adding the same integer to every label rotates the entire environment.
Let $a_s$ be the binary occupancy of site $s$.
For $\boldsymbol{s}=(s_1,\ldots,s_{\nu-1})$, define
\begin{equation}
    d_\nu^S(\boldsymbol{s})
    =
    \sum_{g\in\mathbb Z_N}
    a_g a_{g+s_1}\cdots a_{g+s_{\nu-1}}
\end{equation}
which counts the occupied copies of the oriented pattern $(0,s_1,\ldots,s_{\nu-1})$ as the origin $g$ runs around the ring.
For example, $d_3^S(p,q)$ counts triangles with vertices $(g,g+p,g+q)$.

More generally, $d_\nu^S$ is precisely the cyclic $k$-deck of reconstruction theory, with $k=\nu$.
In particular, Radcliffe and Scott~\cite{radc-scot98jcta} proved that the directed 3-deck determines, up to translation, every subset of $\mathbb Z_p$ when $p$ is prime and almost every subset of $\mathbb Z_N$ as $N\to\infty$.

For the atom-centered environments considered here, however, the offsets $\boldsymbol{s}$ and $-\boldsymbol{s}$ describe the same geometric motif traversed in opposite directions.
Indeed, reversing a planar ring is equivalent in three dimensions to a rotation by $\pi$ about a diameter.
For each $\boldsymbol{s}$, we therefore combine the multiplicities of the two orientations by the pointwise sum
\begin{equation}
    D_\nu^S(\boldsymbol{s})
    =d_\nu^S(\boldsymbol{s})+d_\nu^S(-\boldsymbol{s}).
\end{equation}
This reflected correlation array is redundant: permutations of the offsets and different choices of the reference site can represent the same unlabeled motif, and a motif equivalent to its own reflection can be counted twice. Identifying all entries that represent the same motif and applying the corresponding class-dependent normalization gives the complete reflection-symmetrized deck $D_\nu(S)$, namely the histogram that assigns to every $\nu$-site motif class its number of occurrences. The raw array and this canonical histogram contain the same information, so equality of either is equivalent.

To express the same construction in the atomistic language used above, take the discrete Fourier transform of the occupancy pattern,
\begin{equation}
    \widehat a_m
    \equiv\sum_{s=0}^{N-1}a_s
    \exp\left(-\frac{2\pi\mathrm{i}ms}{N}\right).
\label{eq:ring-fourier-coefficients}
\end{equation}
Because all neighbors have radius $R$ and polar angle $\pi/2$, the density coefficients factorize as
\begin{equation*}
    c_{nlm}=R_n(R)Y_{lm}^*(\pi/2,0)\widehat a_m,
\end{equation*}
where the common cutoff factor has been absorbed into $R_n(R)$. As explicitly shown in the supplementary material,
for $\boldsymbol{m}=(m_1,\ldots,m_{\nu-1})$, the Fourier transform of the deck with respect to its offsets is
\begin{equation}
    \widehat D_\nu^S(\boldsymbol{m})
    =2\operatorname{Re}\!\left[
    \prod_{j=1}^{\nu}\widehat a_{m_j}
    \right],
    \quad
    \text{with}
    \quad
    m_\nu\equiv-\sum_{j=1}^{\nu-1}m_j.
\label{eq:ring-fourier-deck}
\end{equation}
Substituting the factorized coefficients $c_{nlm}$ into a reflection-even invariant and pairing the terms indexed by $\boldsymbol{m}$ and $-\boldsymbol{m}$ gives
\begin{equation}
    B^S_{\nu\boldsymbol n\boldsymbol l}
    =\frac{1}{2}\sum_{\boldsymbol{m}}
    W_{\boldsymbol n\boldsymbol l}(\boldsymbol{m})
    \widehat D_\nu^S(\boldsymbol{m}),
\label{eq:deck-to-atomistic-spectrum}
\end{equation}
where $W_{\boldsymbol n\boldsymbol l}(\boldsymbol{m})$ is determined by the radial basis functions, the spherical harmonics evaluated at $(\pi/2,0)$, and the angular-momentum coupling coefficients; it does not depend on the occupied subset $S$.
Equation~\eqref{eq:deck-to-atomistic-spectrum} shows that equal decks give identical reflection-even $B_{\nu\boldsymbol n\boldsymbol l}$ for every radial and angular channel.
The reflection-odd invariants vanish for these coplanar environments, so their complete order-$\nu$ atomistic spectra are identical.
If the decks agree through order $\nu$, the lower-order spectra and all repeated-index contributions agree as well.
Gr{\"u}nbaum and Moore proved that, for rational-valued cyclic sequences with nonzero total weight, full complex invariants through order six determine the sequence up to translation, with order four sufficient when $N$ is odd~\cite{grun-moor95ac}.
This distinction is important here: Gr{\"u}nbaum and Moore's theorem uses the full complex products, while the reflection-symmetrized deck $D_\nu$ retains only their real parts by identifying opposite orientations.
The lowest-order pair in Fig.~\ref{fig:counterexample-atlas}(a) was already known in music theory: up to rotations and reflections, it is Collins's pair of nine-note collections in an 18-tone pitch space with matching three-note pattern counts~\cite{collins99}.
Building on this idea, Mandereau~\emph{et~al.} later reported two 18-site patterns on a regular 36-gon that are not related by a rotation or reflection, yet have identical decks through $D_4$~\cite{mandereau}.
As atomic environments, they have the same power spectrum, bispectrum, and trispectrum, and are first distinguished by the tetraspectrum.
This planar pair appears in Fig.~\ref{fig:counterexample-atlas}(a) and provides the starting point for the two three-dimensional lifts in panel~(b).
We discuss further the origin of this construction in Sec.~\ref{sec:sound-of-atoms}.

The other pairs in Fig.~\ref{fig:counterexample-atlas}(a) push the same idea to much higher order.
They occupy 54 sites of $\mathbb Z_{154}$, 161 of $\mathbb Z_{462}$, and 9912 of $\mathbb Z_{20790}$, and share all decks through $D_5$, $D_6$, and $D_7$, respectively.
The $D_7$ deck contains approximately $1.86\times10^{24}$ seven-neighbor selections, so direct enumeration is impractical. As detailed in the supplementary material, we instead exploit a unique decomposition into smaller coprime rings, which both makes exact verification tractable and explains why these high-order degeneracies are consistent with Pebody's universal reconstruction bound~\cite{pebo04cpc}: directed decks through order six determine every subset of every finite cyclic group up to translation.
Since none of the constituent patterns used here is a translate of its mirror, each must therefore reveal its orientation by that order.
Their combination can nevertheless remain degenerate under $D_\nu$ until a single correlation detects the orientations of both components at once.
The $\mathbb Z_{20790}$ pair first separates at $D_8$: the two noncongruent environments agree on every cluster of up to seven neighbors (or eight-body order when the central atom is included), providing an explicit counterexample to completeness at $\nu=7$.
These degeneracies are basis-independent, and cannot be resolved irrespective of the resolution at which the correlations are discretized. On the other hand, even for $\nu$ values that should theoretically lift the degeneracy, the very large number of neighbors means that an impractically high angular resolution would be needed to distinguish them using density correlation features, similar to the case discussed in Sec.~\ref{sec:finite-representation}, which, however, relies on a different mechanism.
Further information on the construction and properties of these counterexamples is provided in the supplementary material.

\subsection{Lifting cyclic degeneracies to three dimensions}

The examples in Fig.~\ref{fig:counterexample-atlas}(a) are planar, but their degeneracy is not a consequence of coplanarity.
Figure~\ref{fig:counterexample-atlas}(b) illustrates two lifts using the $\mathbb Z_{36}$ pair, but both constructions are general: if $S,S'\subset\mathbb Z_N$ have identical decks through order $\nu$, either lift produces three-dimensional environments with identical complete $SO(3)$ correlations through the same order.

The first lift adds two identical atoms at the poles of the sphere containing the ring.
Once the polar atoms in a cluster are specified, its remaining geometry is a lower-order cluster of the original ring and therefore has the same multiplicity in both environments.
A reflection of the ring is now realized by a rotation by $\pi$ about a diameter, which exchanges the two poles.
The resulting environments $A$ and $A'$ lie on a single sphere and have identical complete $SO(3)$ correlations through $\nu=4$.

A second lift duplicates the pattern on two parallel rings and allows an arbitrary relative rotation $\alpha$ between them.
Let $\mathcal R_-(S)$ and $\mathcal R_+(S)$ denote the lower and upper rings, respectively. For $S\subset\mathbb Z_N$, define
\begin{equation}
\begin{aligned}
    X_S(\alpha,h)
    &=\mathcal R_-(S)\cup\mathcal R_+(S),\\
    \mathcal R_-(S)
    &=\{\big(R\cos\theta_s,R\sin\theta_s,-h\big):s\in S\},\\
    \mathcal R_+(S)
    &=\{\big(R\cos(\theta_s+\alpha),R\sin(\theta_s+\alpha),h\big):s\in S\},
\end{aligned}
\label{eq:cyclic-lift}
\end{equation}
where $\theta_s=2\pi s/N$.
All neighbors lie on the sphere of radius $\sqrt{R^2+h^2}$.
A rotation or reflection of the planar pattern lifts, respectively, to an axial rotation or to a rotation by $\pi$ that exchanges the two layers.
Because the pattern is copied to both rings, this layer exchange pairs every lifted motif with its reflected counterpart, including motifs that span the two rings.
Equal planar decks through order $\nu$ therefore give equal complete $SO(3)$ correlations through the same order for every $\alpha$ and $h\ne0$. The oriented-deck proof is reported in the supplementary material.

For the $\mathbb Z_{36}$ pair, Eq.~\eqref{eq:cyclic-lift} gives the environments $B$ and $B'$ in Fig.~\ref{fig:counterexample-atlas}(b).
For generic $\alpha$ and $h$ they are noncongruent, but their complete spectra agree through $\nu=4$ and first differ at $\nu=5$.
The lifted pair is therefore a genuinely three-dimensional trispectrum degeneracy that persists when parity-odd $SO(3)$ invariants are retained.

As shown in the supplementary material, the construction extends to any finite collection of mirror-paired rings centered on the same atom. The pair labelled by $p$ may have its own radius, heights $\pm h_p$, and opposite twists $\pm\alpha_p$. A single rotation by $\pi$ exchanges all pairs simultaneously, so the cyclic deck equality is preserved through the same order.
The two-ring lift above is the special case containing only one pair. At the other extreme, one can generate degenerate pairs that are periodic along one direction by stacking an infinite number of lifted rings along the same axis.

\subsection{Finite-representation degeneracies}
\label{sec:finite-representation}

The preceding examples remain degenerate at a fixed correlation order even when the angular basis is complete.
A complementary obstruction arises in every practical representation, for which both the maximum correlation order $\nu_{\max}$ and the angular cutoff $l_{\max}$ are finite.
For any such pair of cutoffs, one can construct a continuous family of noncongruent environments with identical retained features.

Consider two coaxial regular rings containing $q_1$ and $q_2$ atoms, placed at different radii or polar angles so that they cannot be interchanged, and let $\alpha$ be their relative azimuthal rotation.
The sum over a $q_a$-fold ring cancels every spherical-harmonic component except those for which $q_a$ divides $m$.
Consequently, a scalar correlation can depend on $\alpha$ only if it transfers a nonzero total magnetic index between the rings.
Angular-momentum conservation requires this transfer to be divisible by both $q_1$ and $q_2$, and hence to have magnitude at least $\operatorname{lcm}(q_1,q_2)$.

An order-$\nu$ feature with $l\leq l_{\max}$ can transfer at most $\nu l_{\max}$.
For coprime ring sizes, the sufficient condition
\begin{equation}
    q_1q_2=\operatorname{lcm}(q_1,q_2)
    >\nu_{\max}l_{\max}
\label{eq:finite-representation-condition}
\end{equation}
therefore makes every retained invariant independent of $\alpha$, for every radial channel and angular-momentum coupling path.
This spectral invariance holds for every relative rotation.
For generic choices of the two ring geometries, however, the point cloud is congruent to that at $\alpha=0$ only when $\alpha=2\pi k/(q_1q_2)$, with $k\in\mathbb Z$. Every other value changes the cross-ring distances and gives a noncongruent environment with the same retained features.
For any finite $(\nu_{\max},l_{\max})$, one may choose $q_1=N$ and $q_2=N+1$ with $N(N+1)>\nu_{\max}l_{\max}$ (consecutive numbers are always coprime), so a counterexample always exists.
The selection-rule proof of Eq.~\eqref{eq:finite-representation-condition}, together with the noncongruence argument, is given in Sec.~S3 of the supplementary material.

Figure~\ref{fig:counterexample-atlas}(c) uses $(q_1,q_2)=(9,10)$, $\nu_{\max}=8$, and $l_{\max}=10$, so that $90>80$.
All relative angles therefore give the same retained features, while the configurations congruent to $\alpha=0$ occur only at multiples of $360^\circ/90=4^\circ$, shown by the markers in the figure.
The illustrated $2^\circ$ rotation is simply one noncongruent representative: the first possible phase-sensitive correlation occurs at correlation order $19$.
Thus every finite truncation admits a counterexample, although a fixed two-ring pair becomes distinguishable once the cutoffs are increased sufficiently -- even though this cutoff might be so high as to make a model based on the corresponding features numerically unstable and/or computationally impractical.
Although the objects and mechanism differ, this order-by-order failure has a signal-processing analogue: for every prescribed finite autocorrelation order, Yellott and Iverson constructed a different pair of infinite-support band-limited images that agree at that order but differ at the next~\cite{yell-iver92josaa}.

\subsection{Intrinsic rank-three structures}

The lifts in Fig.~\ref{fig:counterexample-atlas}(b) inherit their degeneracy from a planar pattern. Figure~\ref{fig:counterexample-atlas}(d) shows a different class: in each pair the neighbor vectors span all of $\mathbb R^3$, and the degeneracy is intrinsic to the three-dimensional geometry.

For reflection-even $O(3)$ descriptors, the first three pairs, containing 13, 16, and 23 neighbors, have identical complete Gram decks through $D_4$ and first differ at $D_5$. The 13-point construction uses the eight proper rotations of a cube-like twisted square prism: the displayed structures differ only by reflecting one axial point through the origin. The 16-point example uses the twelve proper rotations of a hexagonal bipyramid. Its common core consists of two oppositely twisted hexagons, and the pair changes the orientation of a four-point decoration relative to this core. The 23-point pair uses the full octahedral symmetry group and reorients six points on fixed radial shells around a common unit octahedron. The 26-point construction uses the same octahedral frame but pushes the degeneracy one order higher: its complete Gram decks agree through $D_5$ and first differ at $D_6$.
These equalities do not extend to the complete oriented $SO(3)$ decks. Parity-odd invariants distinguish the 13- and 16-point pairs at order four and the 23- and 26-point pairs at order three -- even though the pairs are \emph{not} enantiomers (for which the parity-odd invariants would change sign). Thus, these examples demonstrate intrinsically three-dimensional degeneracies of reflection-even descriptors, whereas the cyclic lifts above provide the stronger degeneracies that persist when parity-odd channels are retained.

Despite their different appearance, the four pairs follow the same symmetry principle. A common three-dimensional core fixes a reference frame, while a small set of additional points is arranged so that the symmetries of the core redistribute, but do not change, the multiplicities of all cards through the target order. Section~S4 of the supplementary material gives the constructions, coordinates, and proofs in detail.

\section{AI as the ultimate polymath}

Having described the different families of degenerate pairs, we now turn to discussing \emph{how} these counterexamples were found with the assistance of Artificial Intelligence (AI) coding agents, based on Anthropic's Claude (versions Opus 4.8 and Fable) and OpenAI's Codex (versions 5.5 and Sol 5.6).
The strongest results were obtained as part of rather unstructured experiments, in which we prepared a summary of the literature in our field, and we iterated across multiple interactions during which we probed different lines of inquiry, confirming results both manually and using AI-generated code.
We also performed more systematic experiments aimed at determining whether the AI could re-discover the counterexamples from Ref.~\citenum{pozd+20prl} when instructed explicitly to disregard literature published after 2019, and to determine the variability in the outputs of entirely unsupervised searches.

\subsection{The sound of atoms}
\label{sec:sound-of-atoms}

Having used the $\mathbb Z_{36}$ pair above as a technical construction, we now return to its origin.
At the level of note pairs, Mandereau \emph{et al.} made the connection explicit: the musical interval vector is the Patterson function of the corresponding binary pitch-class distribution, making $Z$-relation and homometry equivalent for such collections~\cite{mandereau_zrelation}. In a companion study, the same authors extended this correspondence to larger note patterns and called two collections $Z^{(4)}$-related when every four-note pattern occurs equally often in both. They reported the first such pair in $\mathbb Z_{36}$ whose complete collections are not related by transposition or inversion, which is exactly the pair used here~\cite{mandereau}. As discussed in the supplementary material, the occupied residues represent pitch classes: rotating the cyclic pattern transposes every note by the same amount, whereas reflecting it reverses all interval directions, the operation known as pitch-class inversion. The $k$-vector used in that literature counts how often each $k$-note set class occurs up to these two operations and contains the same information as the reflection-symmetrized cyclic deck used here. Equality of the decks through $D_4$ therefore says that the two collections contain each interval class, three-note set class, and four-note set class with the same multiplicity, even though the complete collections are not related by transposition or inversion.

The significance of this example therefore lies in the translation between fields. Recent studies of AI-assisted mathematics show that language models can contribute both by constructing new arguments and by locating previous solutions obscured in the literature~\cite{feng2026semiautonomousmathematicsdiscoverygemini,balko2026bolzanocasestudiesllmassisted}.
Here, the essential step was to recognize that a musical $Z^{(4)}$-relation and the incompleteness of an atom-centered $\nu=4$ descriptor are the same reconstruction problem expressed in different vocabularies.
In our view, this mode of discovery is far from trivial, and not unlike the way many human discoveries are made~\cite{shi-evan23ncomm}. The over-specialization of modern science is often decried as slowing down discovery, and in this sense LLM assistants simplify tremendously the search for relevant results in far-away disciplines, as well as their translation into a language that is easier to digest by researchers in another domain.

\subsection{Reproducibility of AI searches}

\begin{figure*}[t]
    \centering
    \includegraphics[width=\textwidth]{\arxivpath 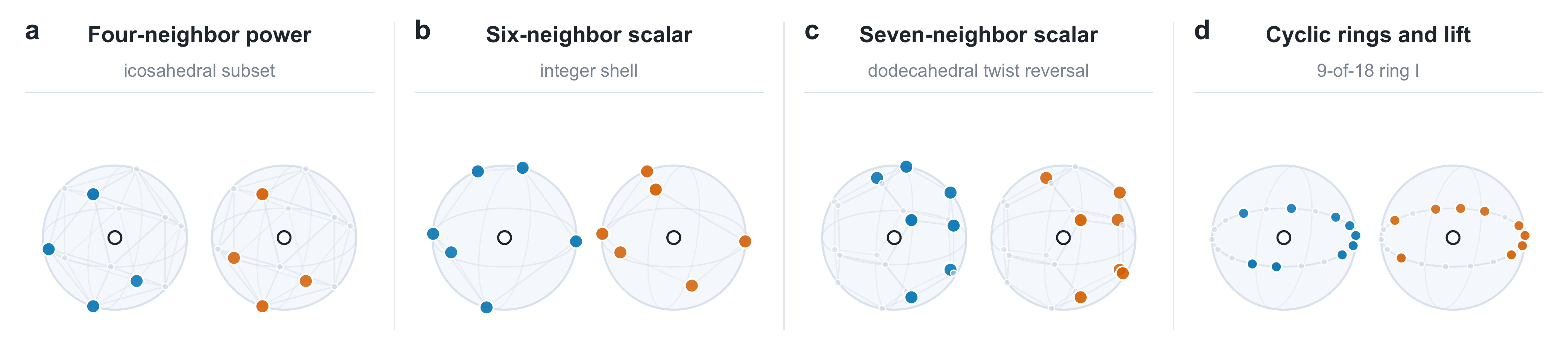}
    \caption{Representative classes recovered by the repeated isolated searches: (a) the four-neighbor power family, represented by an icosahedral subset; (b) the six-neighbor scalar family, represented by an integer shell; (c) the seven-neighbor scalar family, represented by a dodecahedral twist reversal; and (d) the cyclic rings and lift, represented by the $9$-of-$18$ ring I. The scalar degeneracies in panels (b) and (c) are reflection-even, and pseudoscalar channels distinguish the displayed pairs. The blue and orange circles distinguish the two environments in each pair, and the open marker is the central atom. The corresponding parametrizations and the other representatives are reported in the supplementary material.}
    \label{fig:search-class-representatives}
\end{figure*}

\begin{table}[b]
    \centering
    \caption{Classification of the repeated isolated searches. The tuple $(N,d,c)$ gives the neighbor count, family dimension, and codimension defined in the text. The columns headed ``off'' and ``on'' count distinct pairs returned under the offline and online conditions, respectively. The same pair returned in separate runs counts once per run, whereas duplicate outputs within one run count only once. The cyclic row groups constructions with a common deck mechanism rather than a continuous component.}
    \label{tab:search-classification}
    \small
    \setlength{\tabcolsep}{1.2pt}
    \renewcommand{\arraystretch}{1.08}
    \begin{tabular}{@{}p{0.245\columnwidth}ccc@{\hspace{4pt}}ccp{0.235\columnwidth}@{}}
        \toprule
        & & \multicolumn{2}{c}{Codex\ \ } & \multicolumn{2}{c}{Claude} & \\
        \cmidrule{3-6}
        {\raggedright class\par} & $(N,d,c)$ & off & on & off & on & {\raggedright relation\par} \\
        \midrule
        {\raggedright four-neighbor power\par} & $(4,7,2)$ & $8$ & $3$ & $8$ & $7$ & {\raggedright connected family\par} \\
        {\raggedright six-neighbor scalar\par} & $(6,4,11)$ & $1$ & $2$ & $0$ & $0$ & {\raggedright same representative\par} \\
        {\raggedright seven-neighbor scalar\par} & $(7,6,12)$ & $1$ & $1$ & $0$ & $0$ & {\raggedright same representative\par} \\
        {\raggedright cyclic rings and lift\par} & --- & $6$ & $1$ & $2$ & $1$ & {\raggedright same mechanism\par} \\
        \bottomrule
    \end{tabular}
\end{table}

Given the stochastic nature of LLM inference, one may wonder whether each query will find strong counterexamples.  It is also interesting to consider whether models need knowledge of the 2020 Pozdnyakov et al. structures.
To test these questions, we run experiments in which we present the same prompt to Claude Code (Opus v. 4.8) and Codex (5.6 Sol).
We repeat each of two search conditions eight times with each model. Both prompts instruct the model to start from a selection of papers including Refs.~\citenum{kaka12jmiv,bart+13prb,behl11jcp,will+19jcp,drau19prb} and to avoid using information published after 2019. One forbids internet searches, whereas the other encourages searches of older literature in addition to the papers provided (see the supplementary material for the full prompts).

The searches returned many coordinate descriptions, but relabeling, rescaling, or rotating a pair can make the same construction look different. We therefore grouped the outputs by their underlying family. Two pairs belong to the same continuous class if one can be deformed into the other while maintaining deck equality. Figure~\ref{fig:search-class-representatives} shows one representative per class, and Table~\ref{tab:search-classification} reports how often each class was recovered.
The family dimension $d$ measures how many independent changes of shape preserve the degeneracy. After removing global rotations, an $N$-neighbor environment has $D=3N-3$ internal degrees of freedom, and the corresponding codimension is $c=D-d$. A larger $d$ therefore corresponds to a broader family of geometrically distinct but representation-degenerate environments. The first three rows of Table~\ref{tab:search-classification} are continuous classes in this sense. The cyclic examples have different sizes and do not form one continuous family: they are grouped only by their common deck-counting mechanism, so $d$ and $c$ are not defined.

One negative result is particularly revealing. Although the systematic searches were designed in part to test whether the models could rediscover the 2020 counterexamples, none returned the full bispectrum-degenerate pair of Ref.~\citenum{pozd+20prl} and the closest results were the two seven-neighbor dodecahedral twist-reversal reports.
As shown in the supplementary material, if mirror images are treated as equivalent, these examples belong to the same Gram-deck family as the 2020 construction. However, their oriented geometries are different: their parity-even bispectrum components agree, whereas parity-odd, pseudoscalar components distinguish them. The searches therefore recovered a nearby but weaker degeneracy, rather than the complete $SO(3)$-bispectrum counterexample reported in 2020.
It is also remarkable that unsupervised searches, with standardized prompts, led to a large variability of outcomes, and in general to results that were much weaker than those found by interactive ``conversations'' with the LLM. These observations, albeit anecdotal, suggest that defining, and refining, the scientific question at hand remains an important role for the human scientist using AI assistance.

\section{Conclusions}

We have shown examples of atomic structures that cannot be discriminated by fixed-order, permutation-symmetrized atom-centered density correlations, going up to invariant descriptors that are equivalent to knowledge of the full histogram of seven-neighbor (eight-body, including the center) clusters.

These high-order degenerate configurations are very far from any plausible chemical structure, and do little more than reinforce the notion, which is already widely accepted nowadays, that fixed atom-centered invariant descriptors are not the most efficient way of building a machine-learning model of chemical interactions, and add to the list of scenarios in which fixed-order pooling within equivariant architectures limits the expressive power of the resulting models.
What is remarkable is that representation-degenerate structures that are far more powerful than those introduced in Ref.~\citenum{pozd+20prl} were already essentially present in the literature, but in a context so far from that of machine-learning potentials to make them extremely difficult to find.
Given the ease with which LLMs could scout the literature, recognize the relevance of these references, and translate them into results that are directly relevant in this domain, it is hard not to wonder whether the same query, performed in 2019, could have provided in a few hours a definite answer to a research question that required months of human investigation to reach a more restrictive answer.
Our experiments asking the LLMs to ignore literature past 2019 are not conclusive (Ref.~\citenum{pozd+20prl} is almost certainly in the training corpus of all models) but the logic underlying the examples discovered with these prompts is clearly based in prior work, and so it is plausible to answer this question affirmatively.
Even though stronger claims of AI systems being capable of fully autonomous scientific investigation~\cite{Lu_2026} may still be somewhat controversial, the simple usage pattern we exploit here has already a tremendous potential of accelerating discovery.
This is certainly the case when facilitating the translation of results between theoretical fields that speak different dialects of the same universal mathematical language, but it is also likely to be beneficial for empirical disciplines that use characterization and analytical techniques with the potential of being re-purposed to address different research questions than they were designed for, and for which the ability of LLMs to identify patterns across a large corpus of literature can be a remedy for the over-specialization of science.

\section*{Supplementary Material}
The supplementary material contains detailed discussions for each of the classes of problematic structures discussed, as well as the prompts used for the systematic queries, detailed constructions and proofs, validation procedures, the musical interpretation, and complete results and prompts for the repeated searches, and Refs.~\citenum{kurlin2025periodic}--\citenum{deng-mood12cmb} therein.

\begin{acknowledgments}
M.D. and M.C. acknowledge support from an SNSF grant (Project ID: 200020\_214879).
\end{acknowledgments}

\section*{Author Declarations}
\subsection*{Conflict of Interest}
The authors have no conflicts to disclose.

\subsection*{Author Contributions}
\textbf{Michelangelo Domina:} Conceptualization (equal); Data curation (lead); Formal analysis (lead); Software (lead); Validation (lead); Visualization (lead); Writing -- original draft (equal); Writing -- review \& editing (equal).
\textbf{Michele Ceriotti:} Conceptualization (equal); Funding acquisition (lead); Supervision (lead); Validation (supporting); Visualization (supporting); Writing -- original draft (equal); Writing -- review \& editing (equal).

\section*{Data Availability}
The validation scripts, input structures, and interactive Chemiscope~\cite{chor+26joss} visualizations supporting this study will be made openly available on Zenodo~\cite{validation_archive} upon publication of the journal article.

\onecolumngrid
\clearpage


\section*{Supplementary material}

# Supplementary Material for "Using large language models to probe the limits of atom-centered structural descriptors"

Michelangelo Domina[1] and Michele Ceriotti[1,*]

[1]*Laboratory of Computational Science and Modeling, Institut des Matériaux, École Polytechnique Fédérale de Lausanne, 1015 Lausanne, Switzerland*



## S1. CYCLIC COUNTEREXAMPLES OF FIG. 2

As in the main text, $\nu$ denotes the correlation order. An order-$\nu$ atom-centered density correlation is a sum over ordered $\nu$-tuples of neighbors. First consider a term containing $\nu$ distinct neighbors. Their geometry relative to the central atom is completely specified, up to $O(3)$, by the Gram matrix $G_{ab} = \boldsymbol{r}_a \cdot \boldsymbol{r}_b$. Its diagonal entries are the squared neighbor distances, $G_{aa} = r_a^2$, and its off-diagonal entries encode the angles, $G_{ab} = r_a r_b \cos\theta_{ab}$. A $\nu$-card is one such unlabeled $\nu$-neighbor geometry, and the $\nu$-deck is the multiset of all cards obtained by choosing $\nu$ distinct neighbors. These Gram cards describe reflection-even $O(3)$ invariants, whereas the complete $SO(3)$ spectrum distinguishes mirror-related rank-three cards through the signs of scalar triple products, giving what we call oriented cards. Because all scalar triple products vanish for coplanar cards, the Gram and oriented definitions agree in that case. Two decks are therefore equal only if every relevant $\nu$-neighbor geometry occurs with the same multiplicity.

A $\nu$-deck stores each selected $\nu$-neighbor geometry as a separate card. Once the cards are pooled, it no longer records that a particular neighbor may occur in several cards. This loss of correspondence is specific to deck pooling, rather than to invariant point-cloud representations in general. The Simplexwise Centered Distribution introduced by Kurlin [1] organizes the information differently. In three dimensions, it uses two selected points and the centroid as three common references and stores, for every remaining point, one row containing all distances to these references together with a signed quantity that records its side of their plane. Because the measurements in each row remain grouped as properties of the same point, the collection over all selected pairs is complete for finite unlabeled clouds that span all three dimensions, under translations and rotations.

For an $N$-site regular polygon, label the sites $0, \ldots, N-1$ in cyclic order. If a card occupies $s_1 < \cdots < s_\nu$, its cyclic gaps are

$$(s_2 - s_1, \ldots, s_\nu - s_{\nu-1}, N + s_1 - s_\nu). \tag{S1}$$

They count the polygon edges between successive selected sites while going once around the ring. For example, the card $\{0, 2, 5\}$ on an eight-site ring has gaps $(2, 3, 3)$. Choosing a different starting site cyclically permutes this list, while reflection reverses it. In our cyclic-deck validation script [2], we use the lexicographically smallest cyclic permutation of the gap list and its reversal as an exact label for a polygonal card. Identifying a card $S \subset \mathbb{Z}_N$ with the one-periodic sequence $S + N\mathbb{Z}$ turns Eq. (S1) into the consecutive-gap, or time-shift, vector of Ref. [3]. The same canonicalization plays two different roles. In Ref. [3], the vector is built from all points in one period, so it describes the entire periodic sequence and is complete. Here it is built separately for each selected $\nu$-point card. We then retain only the multiset of card labels, which discards the information about which cards overlap in the same sites. Thus, completeness of the full-motif vector does not imply completeness of the $\nu$-deck.

The ordered tuples in a density correlation may repeat a neighbor index. For example, the order-four tuple $(a, a, b, c)$ contains only the three-neighbor card $\{a, b, c\}$, with the additional instruction that $a$ is used twice. For each card, the correlation sum contains every assignment of such repetition patterns to its neighbors. Consequently these contributions are fixed by the lower-order decks. By comparing $D_1, D_2, \ldots, D_\nu$, the script checks both the terms with $\nu$ distinct neighbors and all repeated-index terms entering an order-$\nu$ density correlation.

The implementation uses exact integer arithmetic. The only non-integer coordinates occur in the 16-point example and belong to $\mathbb{Q}(\sqrt{3})$; they are stored as integer pairs $a + b\sqrt{3}$. It verifies every deck from $D_1$ through the order stated in Fig. 2, either by explicit enumeration or by the exact two-axis certificate derived below. The highest matched orders are summarized in the table.

The two largest cyclic examples cannot be checked by explicitly storing every card. The 161-point structure already contains more than $2.2 \times 10^{10}$ six-neighbor selections, while the 9912-point structure contains approximately $1.86 \times 10^{24}$

---

* michele.ceriotti@epfl.ch

| rank | configuration | matched deck | cards | classes | enumerated |
|---|---|---|---|---|---|
| 2 | 9 of 18 cyclic sites | $D_3$ | 84 | 27 | ✓ |
| 2 | 18 of 36 cyclic sites | $D_4$ | 3060 | 828 | ✓ |
| 2 | 54 of 154 cyclic sites | $D_5$ | 3 162 510 | 1 019 738 | ✓ |
| 2 | 161 of 462 cyclic sites | $D_6$ | $\sim 2.201 \times 10^{10}$ | – | × |
| 2 | 9912 of 20790 cyclic sites | $D_7$ | $\sim 1.861 \times 10^{24}$ | – | × |
| 3 | 13-point pair | $D_4$ | 715 | 298 | ✓ |
| 3 | 16-point pair | $D_4$ | 1820 | 394 | ✓ |
| 3 | 23-point pair | $D_4$ | 8855 | 1232 | ✓ |
| 3 | 26-point pair | $D_5$ | 65780 | 19368 | ✓ |

TABLE S1. Highest matched decks for the finite configurations in Fig. 2. The rank is the dimension of the linear span of the neighbor vectors. A checkmark indicates that every card was enumerated and the complete class counters, including multiplicities, were compared directly. A cross indicates that direct enumeration is impractical, so the equality is instead established exactly by the two-axis argument presented below.

seven-neighbor selections. Their equality is instead verified through the directed cyclic counts introduced in the main text. If $a_s \in \{0,1\}$ records whether site $s$ is occupied in $S \subset \mathbb{Z}_N$, the multiplicity of the oriented motif with offsets $\boldsymbol{s} = (s_1, \ldots, s_{\nu-1})$ is

$$d_\nu^S(s_1, \ldots, s_{\nu-1}) = \sum_{g=0}^{N-1} a_g a_{g+s_1} \cdots a_{g+s_{\nu-1}}, \tag{S2}$$

which counts occupied $\nu$-tuples with the prescribed relative offsets. In Fourier space the same counts have a particularly simple form. We use the discrete Fourier transform

$$\widehat{a}_m = \sum_{s=0}^{N-1} a_s \exp\left(-\frac{2\pi \mathrm{i} m s}{N}\right) \tag{S3}$$

and Fourier transform $d_\nu^S$ with respect to each offset $s_j$. To keep the reference point explicit, label it $j = \nu$ and set $s_\nu = 0$. The change of variables $t_j = g + s_j$ is then bijective, with $t_\nu = g$. Finally, define $m_\nu \equiv -\sum_{j=1}^{\nu-1} m_j \pmod N$, so that $\sum_{j=1}^{\nu} m_j \equiv 0 \pmod N$. Writing $\boldsymbol{t} = (t_1, \ldots, t_\nu)$, substitution gives

$$\begin{aligned}
\widehat{d}_\nu^S(m_1, \ldots, m_{\nu-1}) &\equiv \sum_{s_1,\ldots,s_{\nu-1}} d_\nu^S(s_1, \ldots, s_{\nu-1}) \exp\left(-\frac{2\pi\mathrm{i}}{N}\sum_{j=1}^{\nu-1} m_j s_j\right) \\
&= \sum_{\boldsymbol{t}\in\mathbb{Z}_N^\nu} \left(\prod_{j=1}^{\nu} a_{t_j}\right) \exp\left[-\frac{2\pi\mathrm{i}}{N}\sum_{j=1}^{\nu-1} m_j (t_j - t_\nu)\right] \\
&= \sum_{\boldsymbol{t}\in\mathbb{Z}_N^\nu} \left(\prod_{j=1}^{\nu} a_{t_j}\right) \exp\left(-\frac{2\pi\mathrm{i}}{N}\sum_{j=1}^{\nu} m_j t_j\right) \\
&= \prod_{j=1}^{\nu} \widehat{a}_{m_j}.
\end{aligned} \tag{S4}$$

Thus an order-$\nu$ correlation is represented by products of $\nu$ density coefficients whose frequency indices sum to zero. The transform is invertible, so equality of all products in Eq. (S4) is exactly equivalent to equality of all real-space counts in Eq. (S2). Reflection sends $a_s$ to $a_{-s}$ and hence $\widehat{a}_m$ to $\widehat{a}_{-m} = \widehat{a}_m^*$. The pointwise reflection symmetrization of the directed counts is therefore

$$D_\nu^S(\boldsymbol{s}) = d_\nu^S(\boldsymbol{s}) + d_\nu^S(-\boldsymbol{s}), \tag{S5}$$

and its Fourier transform is

$$\widehat{D}_\nu^S(m_1, \ldots, m_{\nu-1}) = 2\,\mathrm{Re}\left[\widehat{a}_{m_1} \cdots \widehat{a}_{m_{\nu-1}} \widehat{a}_{-m_1-\cdots-m_{\nu-1}}\right]. \tag{S6}$$

The array $D^S_\nu(\boldsymbol{s})$ is a redundant encoding of the cyclic motif histogram: permutations of the offsets and different choices of the reference site can represent the same unlabeled motif, and a motif equivalent to its own reflection can be counted twice. Identifying all entries that represent the same motif and applying the corresponding class-dependent normalization gives the complete deck $D_\nu(S)$, the multiset containing every motif class with its number of occurrences. The raw array and the canonical deck contain the same information, so equality of either is equivalent. In finite-group terminology, this is the binary subset deck for the dihedral action on the ring [4].

We now return to the finite construction and document how the three large cyclic pairs in Fig. 2a were generated. In all three cases, the number of ring sites is obtained by $N = n_0 n_1 n_2$, with $n_0, n_1, n_2$ pairwise coprime. The Chinese remainder theorem relabels each site $t \in \mathbb{Z}/N\mathbb{Z}$ on the ring by the unique triple

$$(\eta, u, v) \in \mathbb{Z}/n_0\mathbb{Z} \times \mathbb{Z}/n_1\mathbb{Z} \times \mathbb{Z}/n_2\mathbb{Z}, \tag{S7}$$

where the three cyclic indices have sizes $n_0$, $n_1$, and $n_2$, respectively. We read this relabeling as the coordinate $\eta$ partitioning the ring into $n_0$ disjoint slices, each containing $n_1 n_2$ sites indexed by $(u, v)$. This is only a relabeling of the one-dimensional ring, not an embedding into a three-dimensional space.

On this partition, each example combines two specified small binary patterns, denoted $A$ and $B$, on rings of sizes $n_1$ and $n_2$. We refer to these smaller patterns as *modules*. The parameters and occupied-site patterns used for the three examples are shown in Fig. S1; the occupied residues are also listed explicitly in our cyclic-deck validation script [2].

Under the CRT relabeling, the full-ring occupancy $a_s$ introduced above is written $a_{\eta,u,v}$. Let $\alpha_u = \mathbf{1}_{u\in A}$ and $\beta_v = \mathbf{1}_{v\in B}$ be the occupancy functions of the two modules in that row. For the first member $X$ of the reported pair, the full-ring occupancy is

$$a_{\eta,u,v} = \mathbf{1}_{\eta=0}\, \alpha_u + \mathbf{1}_{\eta\neq 0}\, \beta_v. \tag{S8}$$

In the slice $\eta = 0$, the occupancy is $a_{0,u,v} = \alpha_u$ and does not depend on $v$. Thus, for every occupied site $u \in A$, all $n_2$ points $(0, u, v)$ with $v \in \mathbb{Z}_{n_2}$ are occupied: the pattern $A$ is copied $n_2$ times along the $v$ coordinate. In every slice $\eta \neq 0$, the occupancy is instead $a_{\eta,u,v} = \beta_v$ and does not depend on $u$. Hence every occupied site $v \in B$ is copied at all $n_1$ values of $u$, and this is repeated in each of the $n_0 - 1$ nonzero $\eta$-slices. For the 54-of-154 example, $(n_0, n_1, n_2) = (2, 7, 11)$ and each module contains three occupied sites. The $\eta = 0$ slice therefore contributes $11\times 3 = 33$ sites and the nonzero slice contributes $7 \times 3 = 21$, for a total of 54 occupied sites on a ring of size $2 \times 7 \times 11 = 154$.

The partner $X'$ is obtained by reflecting $A$ while leaving $B$ unchanged. A reflection of a cyclic module has the general form $u \mapsto c - u$: it is the standard reflection $u \mapsto -u$ followed by a cyclic shift by $c$. These shifts can be chosen independently for $A$ and $B$ without changing the full configuration up to rotation. Indeed, for any module shifts $\Delta_A \in \mathbb{Z}_{n_1}$ and $\Delta_B \in \mathbb{Z}_{n_2}$, the Chinese remainder theorem gives a shift $\Delta \in \mathbb{Z}_N$ satisfying

$$\Delta = 0 \pmod{n_0}, \qquad \Delta = \Delta_A \pmod{n_1}, \qquad \Delta = \Delta_B \pmod{n_2}.$$

The global rotation $t \mapsto t + \Delta$ preserves the $\eta$-slices, shifts $A$ by $\Delta_A$, and shifts $B$ by $\Delta_B$. We may therefore choose the origins of the two module rings independently and write the reflection simply as $\alpha_u \mapsto \alpha_{-u}$.

We now determine how this partial reflection can affect a density correlation. As shown above, an order-$\nu$ cyclic correlation is represented in Fourier space by products of $\nu$ density coefficients whose frequencies sum to zero. For the present construction, the repeated copies of $A$ and $B$ make most Fourier coefficients vanish: the nonzero coefficients are confined to two lines in frequency space. Deriving these two lines will allow us to track separately the orientation information carried by the two modules. Take the discrete Fourier transform with respect to all three CRT coordinates,

$$\widehat{a}_{\lambda,\mu,\kappa} = \sum_{\eta=0}^{n_0-1} \sum_{u=0}^{n_1-1} \sum_{v=0}^{n_2-1} a_{\eta,u,v} \exp\left[-2\pi\mathrm{i}\left(\frac{\lambda\eta}{n_0} + \frac{\mu u}{n_1} + \frac{\kappa v}{n_2}\right)\right].$$

Here $\lambda$ is conjugate to the slice coordinate $\eta$, while $\mu$ and $\kappa$ are conjugate to the module coordinates $u$ and $v$, respectively. The transforms of the two modules are

$$\widehat{\alpha}_\mu = \sum_{u=0}^{n_1-1} \alpha_u e^{-2\pi\mathrm{i}\mu u/n_1}, \qquad \widehat{\beta}_\kappa = \sum_{v=0}^{n_2-1} \beta_v e^{-2\pi\mathrm{i}\kappa v/n_2}$$

Consider first the contribution $\mathbf{1}_{\eta=0}\alpha_u$ in Eq. (S8), which encodes module $A$. It is nonzero only at $\eta = 0$ and is independent of $v$. Its Fourier transform therefore contains the finite sum

$$\sum_{v=0}^{n_2-1} e^{-2\pi\mathrm{i}\kappa v/n_2} = \begin{cases} n_2, & \kappa = 0, \\ 0, & \kappa \neq 0. \end{cases}$$

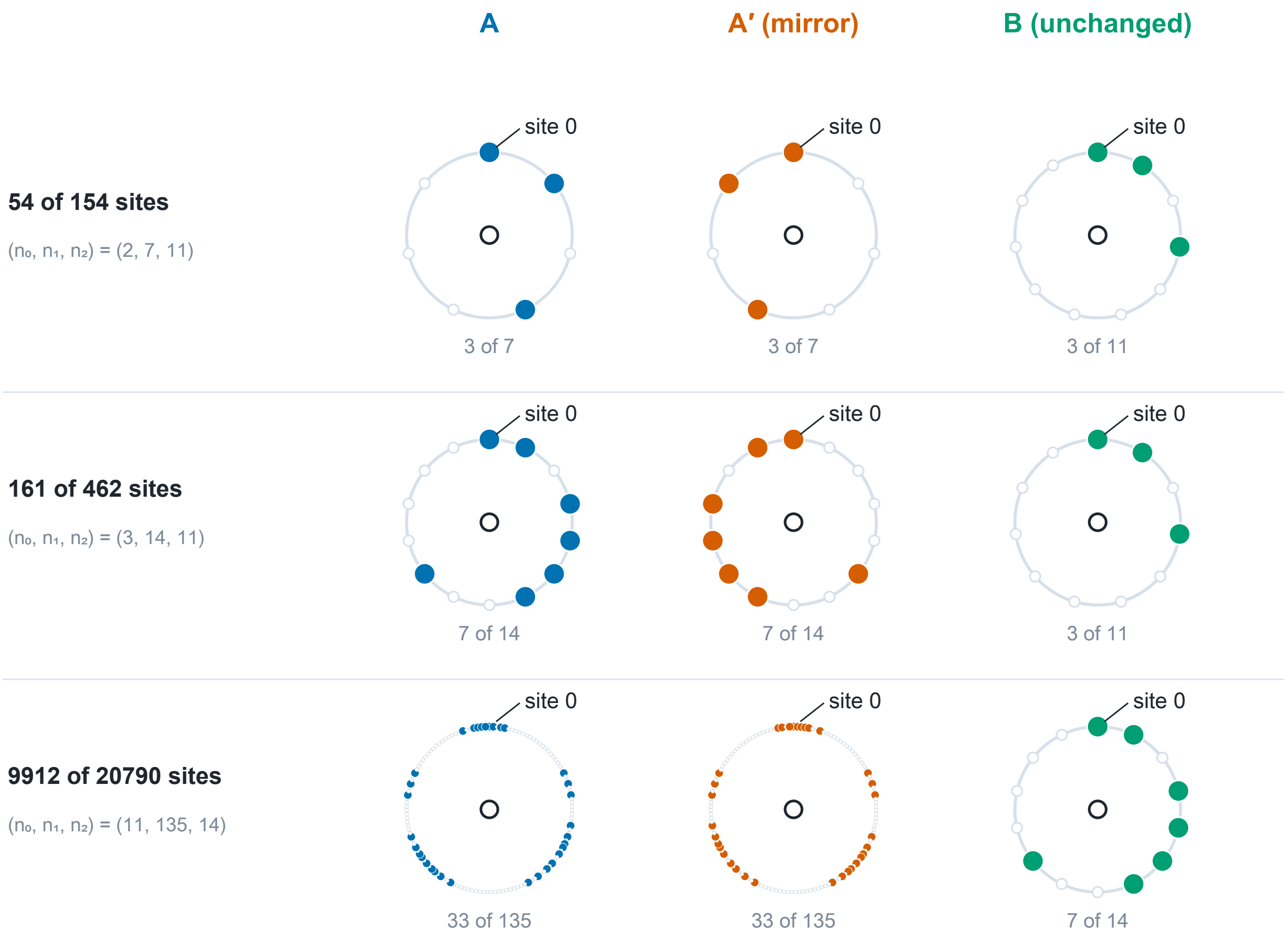


FIG. S1. Cyclic modules used to construct the three large pairs in Fig. 2a. In each row, $A \subset \mathbb{Z}/n_1\mathbb{Z}$ (blue) and $B \subset \mathbb{Z}/n_2\mathbb{Z}$ (green) are the two modules used to build the first configuration $X$. The partner $X'$ replaces $A$ by its mirror $A' = -A$ (orange) and leaves $B$ unchanged. The module origins are chosen so that site 0 lies on the reflection axis. Filled and open markers denote occupied and unoccupied sites.

The contribution from $A$ is therefore $n_2\widehat{\alpha}_\mu \mathbf{1}_{\kappa=0}$.

The second contribution, $\mathbf{1}_{\eta\neq 0}\beta_v$, encodes module $B$ and is independent of $u$. The sum over the $n_1$ values of $u$ gives $n_1\mathbf{1}_{\mu=0}$ by the same argument. Unlike the first contribution, this one is present in every selector slice $\eta = 1, \dots, n_0 - 1$. Summing these copies with their Fourier phases produces the factor

$$h_\lambda \equiv \sum_{\eta=1}^{n_0-1} e^{-2\pi i\lambda\eta/n_0} = \begin{cases} n_0 - 1, & \lambda = 0, \\ -1, & \lambda \neq 0. \end{cases}$$

For $\lambda \neq 0$, the corresponding sum including $\eta = 0$ is zero; omitting that term, which equals one, leaves $h_\lambda = -1$. Combining the two contributions gives

$$\widehat{a}_{\lambda,\mu,\kappa} = n_2\widehat{\alpha}_\mu \, \mathbf{1}_{\kappa=0} + n_1 h_\lambda \widehat{\beta}_\kappa \, \mathbf{1}_{\mu=0}. \tag{S9}$$

For a fixed slice frequency $\lambda$, it is helpful to view $\widehat{a}_{\lambda,\mu,\kappa}$ as a table whose row and column indices are $\mu$ and $\kappa$. The repeated copies of $A$ contribute only along the row $\kappa = 0$, where the values are proportional to $\widehat{\alpha}_\mu$. The repeated copies of $B$ contribute only along the column $\mu = 0$, where the values are proportional to $\widehat{\beta}_\kappa$. The Fourier coefficients therefore form a cross in the $(\mu, \kappa)$ plane: its row and column are the "two axes." Their intersection is

$$\widehat{a}_{\lambda,0,0} = n_2|A| + n_1 h_\lambda |B|,$$

which is real because $h_\lambda$, $|A|$, and $|B|$ are real.

With the module origin chosen as above, the reflected module used in the partner configuration is $A' = \{-u \bmod n_1 : u \in A\}$. Its occupancy function is $\alpha'_u = \alpha_{-u}$, and its Fourier coefficients satisfy

$$\widehat{\alpha}'_\mu = \widehat{\alpha}_{-\mu} = \widehat{\alpha}^*_\mu.$$

Thus, passing from $X$ to $X'$ conjugates every coefficient on the $A$ row. The $B$ column is unchanged because the same module $B$ is used in both configurations, and the intersection is unchanged because it depends on $A$ only through the real number $|A|$.

It remains to determine the first correlation order that can detect this partial reflection. Equation (S6) shows that each Fourier component of the reflection-symmetrized deck is twice the real part of a frequency-conserving product. Choose one such set of $\nu$ frequency triples and denote the corresponding product for $X$ by

$$Q_X = \prod_{j=1}^{\nu} \widehat{a}_{\lambda_j,\mu_j,\kappa_j}.$$

The corresponding Fourier-deck components are

$$\widehat{D}_X = 2\,\mathrm{Re}(Q_X), \qquad \widehat{D}_{X'} = 2\,\mathrm{Re}(Q_{X'}).$$

They can differ only if $\mathrm{Re}(Q_X) \neq \mathrm{Re}(Q_{X'})$.

If $Q_X = 0$, at least one factor in the product is zero. Reflecting $A$ conjugates the coefficients on the $A$ row and leaves the $B$ column unchanged, so it preserves every zero coefficient. The same factor therefore vanishes for $X'$, giving $Q_{X'} = 0$. All zero products already match in Fourier space, so we can focus only on the nonzero products.

For a product to be nonzero, every factor must lie on the Fourier-space cross. We sort these factors according to whether they lie on the $A$ row, on the $B$ column, or at their intersection. The intersection coefficients are real and identical for $X$ and $X'$, as are the factors $n_1$, $n_2$, and $h_{\lambda_j}$. They can therefore be collected into a common real prefactor $C$. If $P_\alpha$ denotes the product of the remaining module coefficients $\widehat{\alpha}_{\mu_j}$ from the $A$ row, and $P_\beta$ the corresponding product of $\widehat{\beta}_{\kappa_j}$ from the $B$ column, then

$$Q_X = C P_\alpha P_\beta.$$

We now use frequency conservation to determine what products $P_\alpha$ and $P_\beta$ can occur. Equation (S4) requires the total frequency of $Q_X$ to vanish. Each factor in $Q_X$ carries the three labels $(\lambda_j, \mu_j, \kappa_j)$, and these labels add when the factors are multiplied. Since the three CRT coordinates are independent, the total frequency vanishes only when it vanishes separately along each coordinate:

$$\sum_{j=1}^{\nu} \lambda_j \equiv 0 \pmod{n_0}, \quad \sum_{j=1}^{\nu} \mu_j \equiv 0 \pmod{n_1}, \quad \sum_{j=1}^{\nu} \kappa_j \equiv 0 \pmod{n_2}.$$

Together with the cross-shaped support, these conditions separate frequency conservation between the two modules. Every factor on the $B$ column has $\mu_j = 0$, so only the $A$-row factors can cancel one another to make the total $\mu$ frequency zero. The frequencies in $P_\alpha$ must therefore sum to zero by themselves. Conversely, every factor on the $A$ row has $\kappa_j = 0$, so the frequencies in $P_\beta$ must also sum to zero by themselves. Thus every valid full-ring product contains a frequency-conserving correlation term of $A$ and one of $B$.

Replacing $A$ by $A'$ conjugates $P_\alpha$, while $P_\beta$ and $C$ remain unchanged. Thus $Q_{X'} = C P^*_\alpha P_\beta$, and the difference between the corresponding deck components is determined by

$$\mathrm{Re}(Q_{X'}) - \mathrm{Re}(Q_X) = 2C\,\mathrm{Im}(P_\alpha)\,\mathrm{Im}(P_\beta). \tag{S10}$$

Equation (S10) shows why the imaginary parts encode orientation. Reflection conjugates $P_\alpha$: its real part is unchanged, whereas its imaginary part changes sign. The full deck component can therefore change only if both $P_\alpha$ and $P_\beta$ have nonzero imaginary parts.

Let $p$ be the first order at which a frequency-conserving product of coefficients from module $A$ can have a nonzero imaginary part, and define $q$ analogously for module $B$. Below order $p$, the Fourier correlations of $A$ contain no orientation-sensitive signal, and below order $q$, the same is true for $B$. For example, if every second-order product from $A$ is real,

$$\widehat{\alpha}_\mu \widehat{\alpha}_{-\mu} = |\widehat{\alpha}_\mu|^2,$$

and if at least one third-order product

$$\widehat{\alpha}_{\mu_1}\widehat{\alpha}_{\mu_2}\widehat{\alpha}_{-\mu_1-\mu_2}$$

is complex, then $p = 3$ and $A$ first reveals its orientation at third order. If instead $p = 4$, then all of its third-order products are real and orientation first appears at fourth order. The role of $q$ is identical for module $B$.

The thresholds also have a direct real-space interpretation. Let $d_r^A(\boldsymbol{s})$ be the number of occurrences of the directed motif $\boldsymbol{s}$ in module $A$, using the definition in Eq. (S2). Reflection reverses every offset, so the mirror $A'$ satisfies

$$d_r^{A'}(\boldsymbol{s}) = d_r^A(-\boldsymbol{s}).$$

Consequently, $p$ is the first order at which some directed motif has different multiplicity in $A$ and $A'$:

$$p = \min\Big\{r : \ d_r^A(\boldsymbol{s}) \neq d_r^{A'}(\boldsymbol{s}) \text{ for some } \boldsymbol{s}\Big\}.$$

Equation (S4) shows that this is equivalent to the Fourier definition above. The Fourier transform of $d_r^A$ is the collection of frequency-conserving products of $r$ module coefficients, and reflection conjugates each product. If every product is real, conjugation leaves the transformed counts unchanged. If instead one is non-real, the transformed counts change and since the Fourier transform is invertible, at least one real-space count must then change as well. In our cyclic-deck validation script [2], we use the real-space criterion and compare exact integer counts to obtain $p$ for $A$ and, analogously, $q$ for $B$.

Finally, consider a full product of order $\nu$. To distinguish $X$ from $X'$, it must contain enough $A$-row factors to make $\mathrm{Im}(P_\alpha) \neq 0$ and enough $B$-column factors to make $\mathrm{Im}(P_\beta) \neq 0$. By definition, this requires at least $p$ factors from $A$ and $q$ from $B$, so $\nu \geq p+q$. Intersection factors are real and provide no orientation information, so they cannot lower this minimum. The reflection-symmetrized decks are therefore identical at every order $\nu \leq p+q-1$.

To see that the decks do differ at order $p+q$, combine the first correlation of $A$ that changes under reflection with the first such correlation of $B$. Minimality of $p$ and $q$ excludes the real zero modes $\widehat{\alpha}_0$ and $\widehat{\beta}_0$ from these products. Since $P_\alpha$ and $P_\beta$ already conserve the $\mu$ and $\kappa$ frequencies separately, setting every $\lambda_j$ to zero makes their combined product conserve all three CRT frequencies. Equation (S9) contributes a factor $n_2$ for each $A$-row term and $n_1(n_0-1)$ for each $B$-column term, so the real prefactor in $Q_X = CP_\alpha P_\beta$ is

$$C = n_2^p\big[n_1(n_0-1)\big]^q > 0.$$

Equation (S10) therefore gives a nonzero difference, and the first differing deck is $D_{p+q}$.

These thresholds also admit general bounds. Pebody's results imply that any binary cyclic module is distinguished from a non-translate of its mirror by directed order six, or by order four when its size is odd [5, 6]. Since $n_1$ and $n_2$ are coprime, at least one is odd and hence $p+q \leq 4+6 = 10$. The two-axis construction can therefore agree through at most $D_9$.

For example, $(p,q) = (3,3)$ means that at least three factors are needed from each module. Using this validation script [2], we obtain this threshold for the 54-of-154 pair, whose decks therefore agree through $D_5$ and first differ at $D_6$. We similarly obtain $(p,q) = (4,3)$ and $(4,4)$ for the 161-of-462 and 9912-of-20790 pairs, whose first differences occur at $D_7$ and $D_8$, respectively. Finally, it constructs each full binary pair and verifies exhaustively that no rotation or reflection relates its two members.

The reported choices of $n_0$ are particular members of larger families. Once $A$, $B$, $n_1$, and $n_2$ are fixed, any $n_0 \geq 2$ coprime to both $n_1$ and $n_2$ gives the same construction on a ring of size $N = n_0n_1n_2$, with $|X| = n_2|A|+(n_0-1)n_1|B|$. The thresholds $p$ and $q$ are properties of the two modules and therefore do not change with $n_0$, so the resulting pair remains deck-degenerate through $D_{p+q-1}$. Changing $n_0$ produces a different finite ring and noncongruence under rotations and reflections should be checked for each chosen value.

The use of cyclic factors to produce correlation ambiguities has a precedent in a different setting. Deng and Moody transferred a finite cyclic ambiguity to distinct infinite aperiodic model sets with identical two- and three-point correlation measures [7], and later obtained weighted six-colour examples agreeing through fifth order [8]. The resulting correlations are long-run directed averages over infinite sets, while the fifth-order examples use non-binary weights. Neither therefore provides a finite binary counterexample at the higher orders reached here.

## S2. THREE-DIMENSIONAL LIFTS OF CYCLIC DECK DEGENERACIES

*Planar input.* Let $S, S' \subset \mathbb{Z}_N$ be two occupied patterns on the same regular $N$-gon, and suppose that their reflection-symmetrized cyclic decks agree at every order $\nu' \leq \nu$. On the polygon, a global shift $s \mapsto s+c$ is a rotation and a

reversal $s \mapsto c - s$ is a reflection of the cyclic ordering. The deck therefore contains the combined multiplicity

$$D^S_{\nu'}(\boldsymbol{s}) = d^S_{\nu'}(\boldsymbol{s}) + d^S_{\nu'}(-\boldsymbol{s}),$$

rather than the two directed counts separately. A planar card carries no additional handedness: all scalar triple products vanish, and its cyclic reflection can also be realized by a proper three-dimensional rotation through $\pi$ about the corresponding axis in the polygon plane.

*Orientation after lifting.* Lifting the points away from this plane requires more care. A three-dimensional card can have a nonzero scalar triple product, so a deck modulo proper rotations can distinguish its two orientations. We write $D^{SO(3)}_{\nu'}$ for this oriented deck. Equality of the planar cyclic decks does not, by itself, imply equality of these stronger $SO(3)$ decks.

The constructions below preserve the degeneracy because the cyclic reflection is accompanied by an exchange of two identical layers. Together these operations are a proper three-dimensional rotation. Because every occupied polygon site is copied to both layers, every lifted occurrence has the reflected, layer-exchanged partner required by this rotation. Their combined multiplicity is precisely $d^S_{\nu'}(\boldsymbol{s}) + d^S_{\nu'}(-\boldsymbol{s})$, the quantity already fixed by the cyclic deck. We first prove this statement for two layers with an arbitrary relative rotation and then extend it to cylindrical stacks.

**Two copies with an arbitrary relative rotation.** Write $\theta_s = 2\pi s/N$. As in the main text, denote the lower and upper lifted rings by $\mathcal{R}_-(S)$ and $\mathcal{R}_+(S)$, respectively. Place the lower copy at height $-h$ and the upper copy at height $+h$, rotated by an arbitrary angle $\alpha$:

$$\begin{aligned} \mathcal{R}_-(s) &= (R\cos\theta_s, R\sin\theta_s, -h)\,, \\ \mathcal{R}_+(s) &= (R\cos(\theta_s+\alpha), R\sin(\theta_s+\alpha), h)\,, \\ \mathcal{R}_\pm(S) &= \{\mathcal{R}_\pm(s) : s \in S\}\,, \\ X_S(\alpha,h) &= \mathcal{R}_-(S) \cup \mathcal{R}_+(S). \end{aligned} \tag{S11}$$

The first two lines give the position of site $s$ on the lower and upper rings. The third line collects these points over all occupied sites $s \in S$, and the last line combines the two rings into the lifted environment. The same values of $R$, $h$, and $\alpha$ are used for $S$ and $S'$, and all atoms in the two layers have the same species.

A shift $s \mapsto s + c$ simply rotates the entire two-ring environment about the $z$ axis by $\theta_c$. Reversing the order of the sites around the polygon also has a direct geometric realization: a rotation by $\pi$ about the horizontal axis whose direction in the $xy$ plane makes an angle $(\theta_c + \alpha)/2$ with the $x$ axis. This half-turn exchanges the upper and lower rings while sending each site $s$ to its reflected position $c - s$. If $\bar{\epsilon}$ denotes the ring opposite to $\epsilon$, then

$$\mathcal{R}_\epsilon(s) \longmapsto \mathcal{R}_{\bar{\epsilon}}(c-s). \tag{S12}$$

This construction works for every relative rotation $\alpha$: changing $\alpha$ only rotates the horizontal axis used for the half-turn. Since the transformation is a proper rotation, a motif and its reflected, layer-exchanged copy represent the same oriented $SO(3)$ card.

To enumerate the lifted $\nu'$-deck, select any $\nu'$ distinct atoms from the union of the two rings. For each selected atom, forget its height but retain the polygon site onto which it projects and the sign $\epsilon \in \{-,+\}$ of its ring. Choose one selected atom as the reference. If its projected site is $g$, the remaining projected sites can be written as

$$g + s_1, \dots, g + s_{\nu'-1},$$

with all indices understood modulo $N$. The offsets need not be distinct. If two selected atoms are the upper and lower copies of the same polygon site, they have the same projected offset but opposite ring signs. The vector $\boldsymbol{\epsilon} = (\epsilon_0, \dots, \epsilon_{\nu'-1})$ therefore retains the information lost by the projection and, together with $\boldsymbol{s} = (s_1, \dots, s_{\nu'-1})$, specifies the lifted selection.

Because the same values of $R$, $h$, and $\alpha$ are used for $S$ and $S'$, the data $(\boldsymbol{s}, \boldsymbol{\epsilon})$ specify the same three-dimensional card geometry in both lifted environments. Only the number of times that card occurs depends on the occupied pattern.

For fixed admissible data $(\boldsymbol{s}, \boldsymbol{\epsilon})$, each reference site $g$ for which $g, g + s_1, \dots, g + s_{\nu'-1}$ are occupied in $S$ gives exactly one lifted selection. Equation (S2) counts these reference sites, so their number is $d^S_{\nu'}(\boldsymbol{s})$. Repeated projected sites require no separate treatment: they only repeat a binary occupancy factor, and $a_t^2 = a_t$ reduces the count automatically to the corresponding lower-order one.

The half-turn in Eq. (S12) gives a second realization of the same $SO(3)$ card. It reflects the planar offsets, $\boldsymbol{s} \mapsto -\boldsymbol{s}$, and moves every selected atom to the opposite ring, $\boldsymbol{\epsilon} \mapsto \bar{\boldsymbol{\epsilon}}$. This reflected realization appears $d^S_{\nu'}(-\boldsymbol{s})$ times. Since the two realizations are related by a proper rotation, the deck counts them together, giving

$$d^S_{\nu'}(\boldsymbol{s}) + d^S_{\nu'}(-\boldsymbol{s}) = D^S_{\nu'}(\boldsymbol{s}). \tag{S13}$$

This is exactly the reflection-symmetrized planar count, which is the same for $S$ and $S'$ by hypothesis. Different labels $(\boldsymbol{s}, \boldsymbol{\epsilon})$ may sometimes describe the same three-dimensional card. The deck then adds their multiplicities, which preserves equality because each contribution already matches between $S$ and $S'$. We therefore obtain

$$D_{\nu'}^{SO(3)}(X_S(\alpha, h)) = D_{\nu'}^{SO(3)}(X_{S'}(\alpha, h)) \qquad \text{for every } \nu' \leq \nu \text{ and every } \alpha. \tag{S14}$$

The equality is for the full oriented deck: it includes cards distinguished by handedness and, equivalently, all parity-odd correlation channels. The $O(3)$-invariant correlations form a subset and are therefore equal as well. Because $\alpha$ was chosen arbitrarily, the equality holds for every angle applied equally to the upper layer of both structures.

**Cylindrical stacks of mirror-paired rings.** The two-ring lift extends to an arbitrary finite collection of ring pairs centered on the same atom. Let $\mathcal{I} = \{1, \ldots, P\}$ label the pairs. For each $p \in \mathcal{I}$, choose a radius $R_p > 0$, a height $h_p > 0$, and an angle $\alpha_p$. Place the two copies of the cyclic pattern at heights $-h_p$ and $+h_p$, with opposite angular offsets $-\alpha_p$ and $+\alpha_p$:

$$\begin{aligned} \mathcal{R}_{p,\epsilon}(s) &= (R_p \cos(\theta_s + \epsilon\alpha_p), R_p \sin(\theta_s + \epsilon\alpha_p), \epsilon h_p)\,, \qquad \epsilon \in \{-1, +1\}, \\ C_S(\mathcal{P}) &= \{\mathcal{R}_{p,\epsilon}(s) : s \in S,\ p \in \mathcal{I},\ \epsilon \in \{-1, +1\}\}\,, \qquad \mathcal{P} = \{(R_p, h_p, \alpha_p)\}_{p \in \mathcal{I}}. \end{aligned} \tag{S15}$$

In practice $C_S(\mathcal{P})$ denotes the full neighbor list of the central atom. Every pair may have its own radius, height, and twist, and different pairs may also carry different atomic species, provided that the two rings within a pair have the same species and the corresponding pair has the same species in $C_S$ and $C_{S'}$. All rings of $C_S$ carry the cyclic pattern $S$, while the corresponding rings of $C_{S'}$ carry $S'$. The two-ring environment in Eq. (S11) is the special case $\mathcal{I} = \{1\}$ and $\alpha_1 = \alpha/2$: a global rotation by $-\alpha/2$ changes its angular offsets from $(0, \alpha)$ to $(-\alpha/2, +\alpha/2)$.

The symmetric placement makes the same half-turn work for every pair. Let $H_c$ be the rotation by $\pi$ about the horizontal axis with azimuth $\theta_c/2$. In cylindrical coordinates it sends $(R, \phi, z)$ to $(R, \theta_c - \phi, -z)$, and therefore

$$H_c\, \mathcal{R}_{p,\epsilon}(s) = \mathcal{R}_{p,-\epsilon}(c - s) \qquad \text{for every } p, \epsilon, s. \tag{S16}$$

This identity is independent of $R_p$, $h_p$, and $\alpha_p$: the half-turn exchanges the two rings of every pair, reverses the cyclic ordering, and leaves the central atom fixed.

Now select any $\nu'$ neighbors from the stack. Retain the pair and ring labels $(p_j, \epsilon_j)$ of each selected atom, and project the atoms onto the original polygon. For fixed labels and projected offsets $\boldsymbol{s}$, the number of placements is $d_{\nu'}^S(\boldsymbol{s})$, exactly as in the two-ring proof, because every ring carries the same occupied pattern $S$. Equation (S16) maps this selection to the same $SO(3)$ card with offsets $-\boldsymbol{s}$ and labels $(p_j, -\epsilon_j)$; the reflected placements occur $d_{\nu'}^S(-\boldsymbol{s})$ times. The combined multiplicity is therefore $D_{\nu'}^S(\boldsymbol{s})$. This argument already includes cards whose atoms come from several different pairs. Since the reflection-symmetrized planar counts agree for $S$ and $S'$, we obtain

$$D_{\nu'}^{SO(3)}(C_S(\mathcal{P})) = D_{\nu'}^{SO(3)}(C_{S'}(\mathcal{P})) \qquad \text{for every } \nu' \leq \nu. \tag{S17}$$

Thus any finite number of mirror-paired rings may be added without lifting the degeneracy.

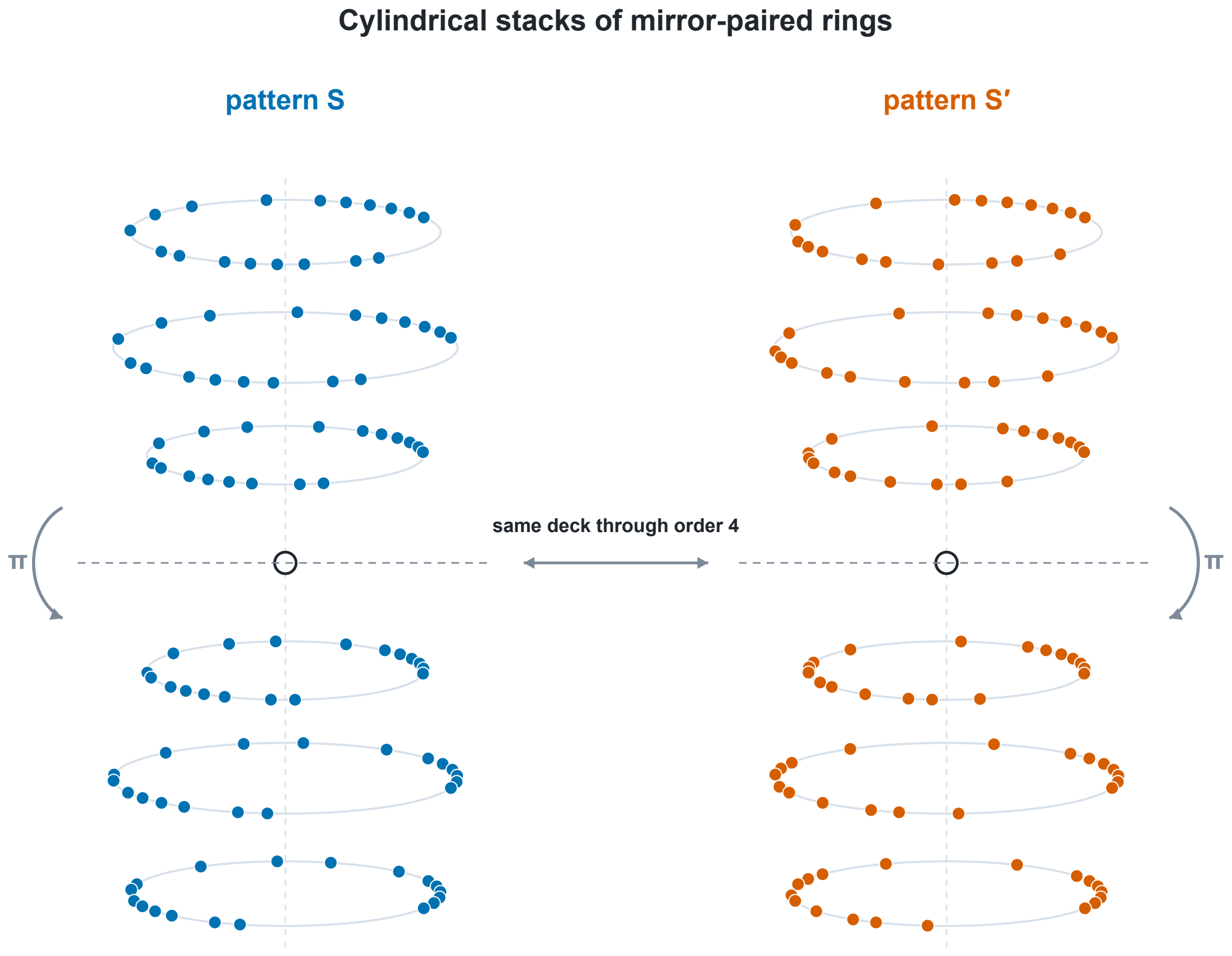


FIG. S2. Finite example of the general cylindrical stack. The index $p$ identifies a pair of rings with its own radius $R_p$, heights $\pm h_p$, and opposite angular offsets $\pm\alpha_p$. Every ring carries $S$ in the left environment and $S'$ in the right one. The central atom is shown once at the common origin. A single rotation by $\pi$ about a horizontal axis exchanges the two rings in every pair and reverses their cyclic ordering, so the two stacks have identical decks through the same order as the original cyclic pair. The construction extends to any finite number of pairs.

## S3. COUNTEREXAMPLES FOR EVERY FINITE SPECTRAL REPRESENTATION

We prove the claim illustrated in Fig. 2c of the main text. Consider a descriptor that retains spherical-harmonic coefficients with $l \le l_{\max}$ and scalar correlations containing at most $\nu_{\max}$ coefficients. For every finite pair $(\nu_{\max}, l_{\max})$, there is a continuous family of noncongruent environments with identical descriptors throughout this finite representation.

**Two-ring construction.** Take two regular rings about the $z$ axis, containing $q_1$ and $q_2$ points, respectively. A $q$-fold ring consists of $q$ equally spaced points and is invariant under a rotation by $2\pi/q$. The two rings have coordinates

$$\boldsymbol{r}_{1a} = \left(R_1 \cos\frac{2\pi a}{q_1}, R_1 \sin\frac{2\pi a}{q_1}, z_1\right), \qquad a = 0, \ldots, q_1 - 1, \tag{S18}$$

$$\boldsymbol{r}_{2b}(\alpha) = \left(R_2 \cos\left(\alpha + \frac{2\pi b}{q_2}\right), R_2 \sin\left(\alpha + \frac{2\pi b}{q_2}\right), z_2\right), \qquad b = 0, \ldots, q_2 - 1. \tag{S19}$$

Here $R_1, R_2 > 0$, and $\alpha$ is the relative azimuthal orientation. The radii and heights are arbitrary; placing the rings at different distances from the central atom prevents them from being interchanged.

For an arbitrary radial basis, the spherical expansion coefficients of this environment can be written as

$$c_{nlm}(\alpha) = \sum_{a=0}^{q_1-1} R_n(r_1) Y^*_{lm}(\theta_1, 2\pi a/q_1) + \sum_{b=0}^{q_2-1} R_n(r_2) Y^*_{lm}(\theta_2, \alpha + 2\pi b/q_2), \tag{S20}$$

where $(r_s, \theta_s)$ specify the radius and polar angle of ring $s$. Weights and cutoff factors can be absorbed into the radial amplitudes.

The effect of the ring symmetry is easiest to see through the azimuthal phase of a spherical harmonic. A rotation about the $z$ axis changes $Y^*_{lm}$ only by a phase,

$$Y^*_{lm}(\theta, \phi) = Y^*_{lm}(\theta, 0) e^{-im\phi}. \tag{S21}$$

For point $b$ on the second ring, this phase is $e^{-im\alpha} e^{-2\pi i m b/q_2}$. The first factor records the orientation of the ring and is common to all its points; the second samples $q_2$ equally spaced phases. More generally, summing these phases over a $q$-fold ring gives

$$\sum_{a=0}^{q-1} \exp\left(-\frac{2\pi i m a}{q}\right) = \begin{cases} q, & q \mid m, \\ 0, & q \nmid m. \end{cases} \tag{S22}$$

The equally spaced phases cancel unless $m$ is a multiple of $q$. We write $q \mid m$ when $m = kq$ for some integer $k$, and $q \nmid m$ otherwise. A $q$-fold ring therefore acts as an angular-frequency filter, retaining only $m = 0, \pm q, \pm 2q, \ldots$. With $\mathbf{1}_{q|m}$ denoting one when $q \mid m$ and zero otherwise, Eq. (S20) becomes

$$c_{nlm}(\alpha) = A^{(1)}_{nlm} \mathbf{1}_{q_1|m} + A^{(2)}_{nlm} \mathbf{1}_{q_2|m} e^{-im\alpha}, \tag{S23}$$

The amplitudes $A^{(s)}_{nlm}$ contain the radial and polar dependence; the explicit phase $e^{-im\alpha}$ is the only dependence on the relative ring orientation. Points on the common axis contribute only to $m = 0$ and do not alter this conclusion.

**Invisibility of the relative rotation.** Consider an arbitrary scalar correlation of order $\nu \le \nu_{\max}$,

$$B_\nu(\alpha) = \sum_{m_1 \ldots m_\nu} C^0_{l_1 m_1 \ldots l_\nu m_\nu} \prod_{a=1}^{\nu} c_{n_a l_a m_a}(\alpha), \qquad l_a \le l_{\max}. \tag{S24}$$

The coupling coefficient can be nonzero only when magnetic angular momentum is conserved,

$$\sum_{a=1}^{\nu} m_a = 0. \tag{S25}$$

Expanding the product in Eq. (S24) assigns each factor to either ring. For one such term, let

$$M_2 = \sum_{a \in I_2} m_a \tag{S26}$$

be the total magnetic index carried by factors from the second ring. Its entire dependence on the relative rotation is $\exp(-i\alpha M_2)$.

The possible values of $M_2$ have a simple Fourier interpretation. Multiplying harmonic coefficients adds their magnetic indices, just as multiplying Fourier modes adds their frequencies. Since every index supplied by the second ring is a multiple of $q_2$, their sum $M_2$ is also a multiple of $q_2$. A scalar feature must have zero total frequency, so the first ring must supply $M_1 = -M_2$. But this sum $M_1$ contains only multiples of $q_1$; consequently, the same nonzero value $|M_2|$ must be divisible by both $q_1$ and $q_2$. The smallest shared frequency is therefore

$$Q = \mathrm{lcm}(q_1, q_2) = q_1 q_2 \tag{S27}$$

when the ring sizes are coprime. No lower frequency can compare their relative phase. For the example in Fig. 2c, the multiples of 9 and 10 first meet at 90.

A finite representation may be unable to construct this shared frequency. Since $|m_a| \le l_a \le l_{\max}$,

$$|M_2| \le \sum_{a \in I_2} |m_a| \le \nu l_{\max} \le \nu_{\max} l_{\max}. \tag{S28}$$

It follows that the sufficient condition

$$q_1 q_2 > \nu_{\max} l_{\max} \tag{S29}$$

excludes every nonzero value of $M_2$: a multiple of $q_1 q_2$ cannot lie strictly between $-q_1 q_2$ and $q_1 q_2$. Every surviving term therefore has $M_2 = 0$, its phase factor equals one, and

$$B_\nu(\alpha) = B_\nu(0) \qquad \text{for all } \nu \le \nu_{\max} \text{ and } l_a \le l_{\max}. \tag{S30}$$

The argument applies independently to every radial channel and every angular momentum coupling path. It proves equality of all $SO(3)$ scalar correlations in the finite box, and therefore also of the inversion-even $O(3)$ subset.

**Non-congruence and existence at every finite cutoff.** The exact geometry does retain the relative phase. In particular, a distance between points on different rings satisfies

$$d_{ab}^2(\alpha) = R_1^2 + R_2^2 + (z_1 - z_2)^2 - 2R_1 R_2 \cos\left(\alpha + \frac{2\pi b}{q_2} - \frac{2\pi a}{q_1}\right). \tag{S31}$$

For generic ring geometries, the configurations at $\alpha = 0$ and $\alpha = \Delta\alpha$ are congruent only at the discrete rotations and reflections generated by their common dihedral symmetry. Choosing

$$\Delta\alpha \notin \frac{2\pi}{q_1 q_2}\mathbb{Z} \tag{S32}$$

therefore gives geometrically distinct environments. All other relative angles, modulo this finite set of identifications and its reflected copies, form a continuous family of noncongruent environments with the same finite descriptor.

For any fixed finite cutoffs, choose $N \ge 3$ with $N(N+1) > \nu_{\max} l_{\max}$ and set $(q_1, q_2) = (N, N+1)$. Consecutive integers are coprime, so Eq. (S29) holds. This produces a counterexample for every finite descriptor truncation, but not a single counterexample for all truncations simultaneously: if $\nu_{\max}$ or $l_{\max}$ is increased, the same pair may eventually be distinguished and the construction must be repeated with larger rings. The result therefore establishes incompleteness at every finite truncation; it does not claim that a fixed environment remains indistinguishable in the limit $\nu_{\max}, l_{\max} \to \infty$.

For Fig. 2c in the main text, we used $q_1 = 9$, $q_2 = 10$, $l_{\max} = 10$, and $\nu_{\max} = 8$. The nine-point ring can contribute only $m = 0, \pm 9$, while the ten-point ring can contribute only $m = 0, \pm 10$. For a scalar feature to detect the relative angle, both rings must contribute a nonzero total angular index. These two totals must sum to zero, because rotating both rings together cannot change a scalar feature. They must therefore be equal and opposite. The smallest possibility is $(M_1, M_2) = (90, -90)$: reaching $M_1 = 90$ requires ten factors with $m = 9$, and reaching $M_2 = -90$ requires nine factors with $m = -10$. This transfer is realized by taking ten $l = m = 9$ factors from the nine-point ring and nine $l = 10, m = -10$ factors from the ten-point ring. The first group couples to $(L, M) = (90, 90)$, while the second couples to $(90, -90)$; these two rank-90 tensors have a nonzero scalar coupling. Thus, for generic ring amplitudes, an order-19 invariant proportional to $\cos(90\alpha)$ exists, and order 19 is the first distinguishing order. In particular, every feature retained in Fig. 2c, for which $\nu \le 8$, is independent of the relative ring angle. Our finite-representation validation script [2] verifies the transfer count by exact enumeration. It also checks that the two environments shown are geometrically distinct: the relative orientations equivalent to $\alpha = 0$ occur at multiples of $360°/\,\mathrm{lcm}(9, 10) = 4°$, whereas the second environment is obtained by rotating the ten-point ring by $2°$.

## S4. INTRINSIC RANK-THREE CONSTRUCTIONS OF FIG. 2D

The four pairs in Fig. 2d are genuinely three-dimensional: their neighbor vectors span $\mathbb{R}^3$, rather than being obtained by lifting a planar pattern. This is what we mean by "rank three." For each pair, two claims must be established. First, every card through the stated order must occur with the same multiplicity in the two environments. Second, the complete environments must not be related by an orthogonal transformation. We keep the construction and verification of these claims separate. We first describe how symmetry was used to design the pairs and then give an independent exact check of the displayed representatives.

**What is counted.** To compare the decks of $X$ and $X'$, we need an exact label for each $\nu'$-neighbor geometry. For selected neighbor vectors $\boldsymbol{r}_1, \ldots, \boldsymbol{r}_{\nu'}$, we use the Gram matrix $G_{ab} = \boldsymbol{r}_a \cdot \boldsymbol{r}_b$. Its diagonal entries give the squared distances from the central atom, while its off-diagonal entries encode the mutual angles. It therefore determines the card up to an orthogonal transformation, as discussed in Sec. S1, and hence captures the reflection-even correlations considered here. An oriented $SO(3)$ card would additionally retain the handedness encoded by scalar triple products. Because the selected neighbors are unlabeled, changing their order only permutes the rows and columns of $G$. We denote the resulting permutation-equivalence class by $\gamma$ and write $D_{\nu',\gamma}(X)$ for its multiplicity in the complete deck $D_{\nu'}(X)$. The label $\gamma$ plays the same role here as the offset vector $\boldsymbol{s}$ in the cyclic construction: both identify a motif whose multiplicity is stored as one entry of the deck. In other words, the complete deck is the multiset containing $D_{\nu',\gamma}(X)$ copies of every card class $\gamma$. Thus $D_{\nu'}(X) = D_{\nu'}(X')$ exactly when they contain the same card classes and $D_{\nu',\gamma}(X) = D_{\nu',\gamma}(X')$ for every $\gamma$.

**Using symmetry to construct equal decks.** Each pair below contains a fixed set of points, which we call the *core*. The remaining points are selected from prescribed symmetry-related sites and form the *decoration*. Two choices of decoration, added to the same core, give the candidate environments $X$ and $X'$.

Directly pairing every occurrence of every card between $X$ and $X'$ can be cumbersome. Instead, let a finite group $G \subset O(3)$ generate the two orbits

$$\mathcal{T}^+ = \{gX : g \in G\}, \qquad \mathcal{T}^- = \{gX' : g \in G\}.$$

We call $\mathcal{T}^+$ and $\mathcal{T}^-$ the two *legs* of the construction. In the constructions below, both legs contain the same number $r$ of distinct environments. Each member $gX$ is a complete alternative environment congruent to $X$, and similarly for $gX'$. Points belonging to different members of a leg are never combined into a larger point cloud.

Every environment within one leg has the same deck because its members are related by elements of the symmetry group $G \subset O(3)$. The purpose of introducing the legs is therefore not to change the deck, but to provide symmetry-related alternatives over which the card counts can be compared. For each card class $\gamma$, we count its occurrences separately in every environment of $\mathcal{T}^+$ and add these multiplicities; we then do the same for $\mathcal{T}^-$. We say that the two legs *balance through order* $\nu$ when these accumulated counts agree for every card class and every order up to $\nu$:

$$\sum_{Y \in \mathcal{T}^+} D_{\nu',\gamma}(Y) = \sum_{Y \in \mathcal{T}^-} D_{\nu',\gamma}(Y) \qquad \text{for every } \nu' \leq \nu \text{ and every } \gamma. \tag{S33}$$

These are bookkeeping sums of card multiplicities over alternative environments, not card counts in a union of their points. Because every member of $\mathcal{T}^+$ is congruent to $X$, and every member of $\mathcal{T}^-$ is congruent to $X'$, Eq. (S33) reduces to

$$r\, D_{\nu',\gamma}(X) = r\, D_{\nu',\gamma}(X'),$$

and cancelling $r$ proves that the representatives have equal decks through order $\nu$.

This also clarifies the limitation of the orbit construction. If Eq. (S33) were checked by exhaustive enumeration, it would be no simpler than comparing $X$ and $X'$ directly: the two sides are just $r$ copies of the multiplicities that we want to compare. The orbit becomes useful only when symmetry gives a simpler way to establish the accumulated equality, for example by pairing whole families of card occurrences between the two orbits. It can then expose the mechanism behind the equality and guide the search for suitable decorations. The first example below provides such a pairing. For every example, our rank-three validation script [2] gives a separate certificate: it selects the cards directly from the two displayed environments, assigns each card its canonical Gram-matrix class, and compares the resulting multiplicities. This calculation uses neither the additional orbit members nor the pairing argument.

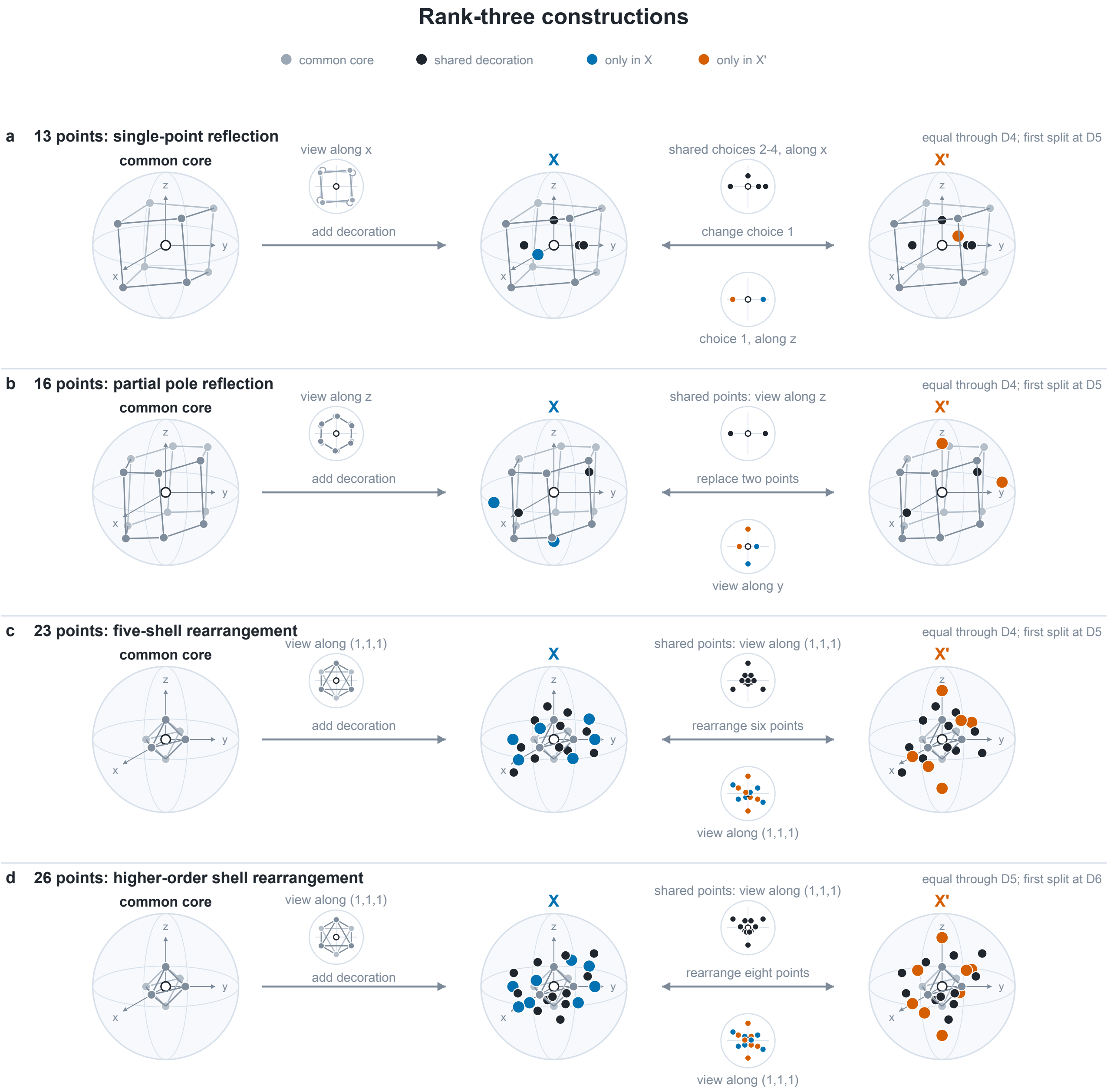


FIG. S3. Construction of the four intrinsic rank-three pairs in Fig. 2d. Panel a: the 13-point single-point reflection. The black points are the shared choices $U_2^+$, $U_3^+$, and $U_4^+$; in particular, $U_4^+$ is the opposite pair. The blue and orange singletons are $U_1^+$ and $U_1^-$. Panel b: the 16-point partial pole reflection, in which two decoration points are replaced. Panel c: the 23-point five-shell rearrangement, in which six points change. Panel d: the 26-point higher-order shell rearrangement, in which eight points change. Each panel shows the common core, one representative $X$, and its partner $X'$. Gray points belong to the core, black points are shared decorations, and blue and orange points occur only in $X$ and $X'$, respectively. The upper insets expose the core and shared decoration from the indicated direction, while the lower inset overlays the points that differ. Radial shell separations are compressed for visibility. Panels a–c have equal decks through $D_4$ and differ at $D_5$; panel d has equal decks through $D_5$ and differs at $D_6$.

The 13-point pair is the simplest example, so we present it in detail to make the group-orbit mechanism explicit and show how card occurrences balance between its two legs. The later examples follow the same general construction, combining a symmetry-invariant core with alternative decorations, but use different symmetry groups and decoration pairings. We therefore describe their specific geometry more compactly rather than repeating the full argument.

**Single-point reflection: 13 points.** The two environments are built in two stages. First, both contain the same eight-point core, shown in gray in the first row of Fig. S3. Four of its points form a square in the plane $x = +7$, and the other four form a congruent square in the plane $x = -7$. The two squares are rotated relative to one another about the $x$ axis, producing a twisted square prism. We denote this common core by $K_8$.

The rotations in $G_{\text{ax}}$ are most easily described through their action on the coordinate directions. Choose any subset of $x$, $y$, and $z$ and reverse the signs of those directions. If the number of reversals is odd, also exchange the two transverse directions $y$ and $z$ to restore the handedness of the frame and make the transformation a proper rotation. The $2^3 = 8$ possible choices of signs give the eight elements of $G_{\text{ax}}$. We take the point with Cartesian coordinates $(x, y, z) = (7, 11, 13)$ as a seed and apply all eight rotations to it:

$$K_8 = G_{\text{ax}}(7, 11, 13). \tag{S34}$$

Thus, $(7, 11, 13)$ is one vertex of the core, and the notation on the right denotes its complete orbit under $G_{\text{ax}}$. Applying any further rotation $g \in G_{\text{ax}}$ only permutes these eight vertices, so

$$gK_8 = K_8.$$

Second, we add a five-point decoration, specified by four binary choices. The first three choices contribute one point each: the point lies at one of the two ends of the $x$, $y$, or $z$ axis. The fourth choice contributes two points: an opposite pair on either the $y$ axis or the $z$ axis. Denoting the two alternatives for choice $j$ by $U_j^+$ and $U_j^-$ gives

$$\begin{aligned} &U_1^\pm = \{(\pm 2, 0, 0)\}, \qquad U_2^\pm = \{(0, \pm 3, 0)\}, \qquad U_3^\pm = \{(0, 0, \pm 3)\}, \\ &U_4^+ = \{(0, 5, 0), (0, -5, 0)\}, \qquad U_4^- = \{(0, 0, 5), (0, 0, -5)\}. \end{aligned} \tag{S35}$$

There are $2^4 = 16$ possible decorations. To label them, choose one symbol $s_j \in \{+, -\}$ for each binary choice $j$. We call the resulting four-symbol label $\boldsymbol{s} = (s_1, s_2, s_3, s_4)$ a sign word. Its $j$th symbol selects the corresponding set $U_j^{s_j}$, so

$$X_{\boldsymbol{s}} = K_8 \cup U_1^{s_1} \cup U_2^{s_2} \cup U_3^{s_3} \cup U_4^{s_4}.$$

Every $X_{\boldsymbol{s}}$ contains the eight core points, three selected singletons, and one selected opposite pair, for 13 points in total. When products of the signs are used below, the symbols $+$ and $-$ are identified with the numbers $+1$ and $-1$.

The rotations in $G_{\text{ax}}$ act on the decoration in exactly the same way as they act on the coordinate axes: they may reverse the selected endpoints, exchange the roles of the $y$ and $z$ choices, and interchange $U_4^+$ with $U_4^-$. Let $g\boldsymbol{s}$ denote the sign word obtained by applying $g$ to the decoration selected by $\boldsymbol{s}$. Since the same rotation leaves the core invariant as a set,

$$gX_{\boldsymbol{s}} = X_{g\boldsymbol{s}}.$$

The transformations between sign words within a symmetry orbit are therefore actual rotations of the complete environment, while the common core remains unchanged.

The pair displayed in Fig. S3 (panel a) is obtained by changing only the first choice:

$$X = X_{(+,+,+,+)}, \qquad X' = X_{(-,+,+,+)}.$$

The choices $U_2^+$, $U_3^+$, and $U_4^+$ are common to both environments and are shown in black. The two environments therefore share twelve points and differ by the reflection of this single point through the origin.

The other 14 sign choices are introduced only to explain why the $D_4$ decks of $X$ and $X'$ agree. The action of $G_{\text{ax}}$ described above changes an even number of signs, after accounting for any exchange of the $y$ and $z$ choices. It therefore preserves the product $s_1 s_2 s_3 s_4$. The eight rotations connect all eight sign words with product $+1$ to one another, and separately connect all eight words with product $-1$. These are the two symmetry orbits. All environments in a given orbit are congruent and therefore have the same deck. Here $X$ and $X'$ are representatives of the two different orbits.

Now select at most four atoms to form a card in some $X_{\boldsymbol{s}}$. Suppose, for example, that the selected atoms come from the core and from $U_1$, $U_2$, and $U_3$, but not from $U_4$. Replacing $U_4^+$ by $U_4^-$, or vice versa, changes the parity of the sign word and moves the environment to the other orbit. This single sign change is a pairing between the two orbits, not a rotation in $G_{\text{ax}}$. Nevertheless, it does not move any atom belonging to the selected card, because that

card contains no atom from $U_4$. These selections therefore consist of exactly the same points. The same argument applies whenever any block $U_j$ is absent from the card. Every four-card that contains a core point necessarily omits at least one of the four decoration blocks and is covered by this pairing. The only remaining possibility is a four-card containing one atom from each of $U_1, U_2, U_3, U_4$ and no core point. The selected atom from $U_4$ is collinear with either the $U_2$ or the $U_3$ atom, and these two cases are related by a quarter turn about the $x$ axis. The two endpoints in $U_4$ produce the parallel and antiparallel geometries once each, while the $U_1$ atom is orthogonal to both. Thus every sign word gives the same two card classes with the same multiplicities. Together, the two pairings establish equality through $D_4$ without listing all $\binom{13}{4} = 715$ selections.

At order five, the situation changes. A card can contain one core point together with one atom from each of the four decoration blocks. No block is then unused: changing any sign changes a point that belongs to the card and the pairing argument stops holding at this order.

**Partial pole reflection: 16 points.** The second example uses the proper rotational symmetries of a hexagonal bipyramid. Let

$$H = \langle R_z(\pi/3), R_x(\pi) \rangle$$

be this order-12 group. Applying its rotations to $(2/3, 1/3, 2/3)$ produces the common 12-point core

$$K_{12} = H(2/3, 1/3, 2/3). \tag{S36}$$

Geometrically, $K_{12}$ consists of two oppositely twisted hexagons. The possible decoration sites are the six equatorial points $e_j = (\cos(j\pi/3), \sin(j\pi/3), 0)$, with $j = 0, \ldots, 5$, and the north and south poles $n = (0, 0, 1)$ and $s = (0, 0, -1)$. The two displayed environments add four of these sites:

$$X = K_{12} \cup \{e_0, e_3, s, e_5\}, \qquad X' = K_{12} \cup \{n, e_0, e_3, e_2\}. \tag{S37}$$

Applying all twelve rotations in $H$ to $X$ gives one leg, and doing the same to $X'$ gives the other. Their summed card counts agree through $D_4$. The undecorated hexagonal bipyramid has a reflection that exchanges the two decorations, but that reflection also reverses the handedness of the common core. It therefore does not map the complete environment $X$ to $X'$. Five-card geometries are the first to detect the different orientation of the decoration relative to the core.

**Common octahedral frame.** The 23- and 26-point pairs use the full order-48 group $G_{\mathrm{oct}}$ of signed coordinate permutations. Both environments contain the unit octahedron

$$K_6 = \{\pm e_x, \pm e_y, \pm e_z\}. \tag{S38}$$

These are the only points at unit distance from the central atom. Any congruence between the complete environments must therefore map this octahedron to itself and must belong to $G_{\mathrm{oct}}$. The core consequently fixes the coordinate frame, while the remaining allowed sites are organized into octahedral orbits. For example, $G_{\mathrm{oct}}(0, 2, 2)$ consists of the 12 distinct signed permutations of $(0, 2, 2)$: there are three possible positions for the zero coordinate and four sign choices for the two nonzero coordinates. Every point in this orbit lies on the shell $r^2 = 8$. The group-orbit search selects subsets of this and other allowed orbits so that the card counts balance between the two legs.

**Five-shell rearrangement: 23 points.** Besides the unit octahedron, the allowed sites come from the orbits represented by $(0, 2, 2)$, $(5, 5, 5)$, $(0, 0, 9)$, and $(0, 0, 14)$, so the complete environment occupies five distinct radial shells. A compact but complete specification of $X$ is

$$\begin{aligned} X = K_6 \cup \{&(-2, 0, \pm 2), (-2, 2, 0), (0, -2, \pm 2), (0, 2, -2), (2, -2, 0), (2, 0, 2), (2, 2, 0),\\ &(-5, -5, 5), (-5, 5, -5), (5, -5, -5), (5, 5, 5), \pm 9e_y, \pm 14e_x\}. \end{aligned}$$

Each occurrence of $\pm$ includes both sign choices; for example, $(-2, 0, \pm 2)$ denotes $(-2, 0, -2)$ and $(-2, 0, 2)$. The display contains 17 noncore points, which together with the six points in $K_6$ give all 23 points of $X$. To obtain $X'$, we leave the six core points and eleven of the noncore points unchanged, and make three replacements:

$$\begin{aligned} \{\pm 14e_x\} &\longrightarrow \{\pm 14e_z\},\\ \{\pm 9e_y\} &\longrightarrow \{\pm 9e_x\},\\ \{(0, 2, -2), (2, 0, 2)\} &\longrightarrow \{(0, 2, 2), (2, 0, -2)\}. \end{aligned}$$

The first two lines move opposite pairs of points to different coordinate axes, while the third changes the orientation of two points on the $r^2 = 8$ shell. In total, six points are removed and six are inserted, so both environments contain

23 points. These changes must occur together: considered jointly, their contributions cancel in every card through $D_4$. Applying $G_{\rm oct}$ produces two 48-environment legs satisfying Eq. (S33). The cancellation fails at $D_5$.

**Higher-order shell rearrangement: 26 points.** The final pair uses the same unit, body-diagonal, and axial shells, but replaces the orbit generated by $(0, 2, 2)$ by the orbit of $(0, 2, 3)$. Here

$$X = K_6 \cup \{(-3, 0, 2), (-3, 2, 0), (-2, 0, \pm 3), (0, -3, -2), (0, -2, \pm 3), (0, 3, -2), \\ (2, \pm 3, 0), (3, 0, 2), (3, 2, 0), (-5, -5, -5), (-5, 5, 5), (5, -5, 5), (5, 5, -5), \pm 9e_y, \pm 14e_x\}.$$

The display contains 20 noncore points, which together with $K_6$ give all 26 points of $X$. To obtain $X'$, we leave the six core points and twelve of the noncore points unchanged, and make the replacements

$$\{\pm 14e_x\} \longrightarrow \{\pm 14e_z\}, \\ \{\pm 9e_y\} \longrightarrow \{\pm 9e_x\}, \\ \{(-3, 0, 2), (0, -3, -2), (0, 3, -2), (3, 0, 2)\} \longrightarrow \{(-3, 0, -2), (0, -3, 2), (0, 3, 2), (3, 0, -2)\}.$$

The axial rearrangement is the same as in the 23-point pair, while four points on the $r^2 = 13$ shell reverse their orientation relative to the octahedral frame. Here the orbit search balances every card through $D_5$, again using two legs of 48 environments. The first difference occurs at $D_6$.

**Direct exact verification.** Our rank-three validation script [2] compares the four displayed pairs directly, without constructing $\mathcal{T}^{\pm}$ or using Eq. (S33).

For every selected set of $\nu'$ neighbors, the script constructs its Gram matrix. Any equivalence between two such matrices must match equal diagonal entries, which are the squared neighbor radii. The script therefore fixes a deterministic order of the exact radial-shell labels, considers all simultaneous row and column permutations within each shell, and retains the lexicographically smallest upper triangle. Counting the resulting canonical matrices gives the complete unlabeled deck, including every multiplicity. For the 16-point pair, all coordinates are multiplied by six to clear denominators; this common rescaling does not affect rank, congruence, deck equality, or the order of the first difference. It also constructs the complete $D_5$ decks of the 13-, 16-, and 23-point pairs and verifies that they differ. For the 26-point pair, equality through $D_5$ is again checked by exhaustive deck enumeration. To establish the first difference at $D_6$, it is enough to exhibit one six-card whose multiplicity changes. The script uses the card

$$W = \{(-5, -5, 5), (-3, -2, 0), (0, -9, 0), (0, 0, -14), (-2, -3, 0), (-3, 0, 2)\}. \tag{S39}$$

It counts every six-card with the same radial-shell composition as $W$ and compares the full canonical Gram matrix. Exactly one copy occurs in $X$ and none in $X'$. The $D_6$ decks are therefore different, and equality through $D_5$ shows that order six is the first split.

## S5. MUSICAL INTERPRETATION OF CYCLIC DECKS

Consider the familiar 12-tone equal-tempered pitch-class system. Notes separated by complete octaves are identified, so, for example, every C receives the same label 0. The twelve pitch classes can be labelled explicitly as follows; the sharp and flat names in the same entry are equivalent in this representation.

| residue | pitch class | residue | pitch class |
|---|---|---|---|
| 0 | C | 6 | F♯/G♭ |
| 1 | C♯/D♭ | 7 | G |
| 2 | D | 8 | G♯/A♭ |
| 3 | D♯/E♭ | 9 | A |
| 4 | E | 10 | A♯/B♭ |
| 5 | F | 11 | B |

The C-major chord consists of C, E, and G and is therefore represented by the subset

$$S = \{0, 4, 7\} \subset \mathbb{Z}_{12}.$$

Placing the twelve labels around a circle and marking the sites 0, 4, and 7 gives exactly the occupied-polygon representation used in Sec. S1. At this level, a pitch-class set records only whether each class is present; it discards

register, multiplicity, voicing, temporal order, duration, and dynamics. Nothing in the construction depends on having twelve pitch classes. Dividing the octave into $N$ equal classes instead gives the cyclic group $\mathbb{Z}_N$, and any such pitch-class set is represented by an occupied subset $S \subset \mathbb{Z}_N$. The larger values of $N$ in our examples can therefore be read as generalized pitch-class spaces rather than as conventional twelve-tone chords.

The relevance of rotations and reflections is not merely geometric: these are standard equivalences in musical set theory. Rotating the C-major pattern by two sites gives

$$T_2\{0,4,7\} = \{2,6,9\},$$

which the table identifies as D, F$^\sharp$, and A: the D-major chord. The absolute notes and the key have changed, but every interval within the chord is unchanged. Transposition therefore allows music theory to recognize the same chord or scale type independently of the key in which it appears.

Reflection captures a different but equally useful relation. All pitch-class labels are understood modulo 12, so choosing the reflection $s \mapsto 7 - s \pmod{12}$ gives

$$I_7(\{0,4,7\}) = \{0,3,7\},$$

which maps C major to C minor. This operation, called *pitch-class inversion*, reverses the direction of every interval while preserving its undirected size. Major and minor triads are therefore distinct chords but have the same interval classes, 3, 4, and 5. Here direction is ignored, so the seven-semitone perfect fifth belongs to interval class 5, its shorter complement around the twelve-tone circle. Treating the triads as members of one set class is useful when the object of the analysis is this intervallic structure rather than key, tonal function, or the ordering of the notes. Pitch-class inversion should not be confused with a chord inversion, which only changes which note is placed in the bass.

More generally, on an $N$-class pitch circle, transposition by $t$ and pitch-class inversion about the axis specified by $c$ act as

$$T_t(S) = \{s + t \pmod{N} : s \in S\}, \qquad I_c(S) = \{c - s \pmod{N} : s \in S\}.$$

These are precisely a rotation and a reflection of the occupied polygon. Quotienting the cyclic cards by the same two operations therefore has a direct musical meaning: a card records a pattern of notes independently of its absolute transposition and of whether its interval pattern is read in its original or mirrored orientation.

Selecting every group of $\nu$ notes from $S$ and sorting these groups by their set class produces a list of how many times each $\nu$-note pattern occurs in $S$. For $\nu = 2$, it records how often each undirected interval class occurs; music theory calls this the interval-class vector, often shortened to interval vector. For $\nu = 3$ it counts three-note patterns, and increasing $\nu$ considers progressively larger groups of notes. The complete list is called the $\nu$-vector in the musical literature [9]. For binary pitch-class sets, it contains the same information as the canonical $\nu$-deck up to reflection; the raw reflected correlation array $D^S_\nu(\boldsymbol{s})$ above is a redundant encoding and can count reflection-symmetric motif classes twice. We write two collections as $Z^{(\nu)}$-related when all of these counts agree, and call the relation nontrivial when one complete collection cannot be obtained from the other by transposition or inversion. The parenthesized superscript distinguishes the order of the relation from the cyclic group $\mathbb{Z}_N$. In musical set theory, inequivalent sets with the same interval-class vector are conventionally called $Z$-related [10], corresponding here to the special case $Z^{(2)}$. For two sets that each contain at least $\nu$ notes, equality of their $\nu$-vectors also implies equality of every lower-order vector [9]. Every pair considered here satisfies this condition. This is exactly the nesting property of the cyclic decks.

The 18-of-36 pair used in the main text is not related by transposition or inversion, yet each interval class, three-note pattern, and four-note pattern occurs equally often in both [9]. Its difference lies only in how these common pieces are assembled into the complete collection, and cannot be recovered by examining at most four notes at a time. In the notation of Ref. [9], this particular pair also satisfies $B = I_{35}(\mathbb{Z}_{36} \setminus A)$: the second set is an inversion of the complement of the first. This does not make the relation trivial here, because the set classes used above identify only transposition and inversion, not complementation.

Previous work established nontrivial $Z^{(3)}$-related sets in $\mathbb{Z}_{18}$ and the first $Z^{(4)}$-related pair in $\mathbb{Z}_{36}$ [9, 11]. The larger cyclic pairs constructed here extend this sequence to $Z^{(5)}$, $Z^{(6)}$, and $Z^{(7)}$ relations. In musical terms, for every pitch-class set type of cardinality at most seven, the two members of the largest pair have the same multiplicity, while the complete collections are not related by transposition or inversion. Their difference first appears in the eight-note content.

This correspondence is useful in both directions. In computer-assisted composition, such pairs could in principle provide a combinatorial device for changing the overall collection of pitches while preserving every smaller note pattern up to a chosen size. In atomistic machine learning, they expose the precise information lost by a descriptor that counts motifs only up to a fixed size: even exact counts of every small motif need not determine the structure that contains them. Musical set theory and atom-centered representations are consequently not just analogous applications: on a cyclic domain, they are two interpretations of the same reconstruction problem.

## S6. EQUIVALENCE CLASSES IN THE REPEATED ISOLATED SEARCHES

The repeated isolated searches returned many pairs that look unrelated in their original coordinates. We classify these results by the family of equal pairs to which they belong, while keeping the Codex and Claude provenances separate. The provenance count records each distinct pair returned within a run. Rediscoveries in separate runs are retained once per run, whereas duplicate outputs of the same pair within one run count only once.

Three relations must be distinguished. Two reports describe the *same representative* if their coordinates agree up to point permutations, exchange of the two partners, a common scale, and independent orthogonal transformations of the partners. Two different representatives belong to the *same equal-pair component* if they can be connected by a path

$$t \longmapsto \big(X(t), Y(t)\big), \qquad D_\nu\big(X(t)\big) = D_\nu\big(Y(t)\big), \tag{S40}$$

along which the points remain distinct and the two environments remain noncongruent. The common deck is allowed to change with $t$. This is weaker than a path inside one fixed descriptor fibre, for which $D_\nu(X(t))$ itself would be constant. Finally, constructions may share a counting mechanism without lying in the same continuous configuration space. This last relation is the appropriate one for cyclic pairs with different ring sizes or different numbers of neighbors.

For an environment with $N$ neighbors, the configuration space modulo rotations has dimension

$$D = 3N - 3. \tag{S41}$$

If an equal-pair family has dimension $d$ after quotienting independent rotations of its two members, we define its codimension as $c = D - d$. We determine $d$ in two complementary ways. First, an explicit parametrization supplies $d$ independent shape parameters for which the two decks remain exactly equal. This proves that the family has dimension at least $d$.

To show that no additional shape parameters have been missed, we use the Gram matrices of the cards. Once the correspondence between cards, and between the points within each card, has been fixed, equality of each matched pair of cards is written as equality of the corresponding Gram-matrix entries. We collect all scalar differences between these matched entries into a vector $\boldsymbol{F}(X, Y)$, so the deck-equality equations for that correspondence are $\boldsymbol{F}(X, Y) = 0$. The Jacobian used below is

$$J = \frac{\partial \boldsymbol{F}}{\partial(X, Y)},$$

the derivative of these Gram-entry differences with respect to all $6N$ Cartesian coordinates of the two environments. At a generic regular solution, rank $r$ means that the Gram equalities impose $r$ independent local conditions, leaving a solution set of dimension $6N - r$. Three dimensions in each environment correspond only to rigid rotation and do not change its shape. Removing these six directions leaves $6N - r - 6$ independent shape parameters. If this number equals the number supplied by the explicit parametrization, the value of $d$ is exact. At a highly symmetric representative, several Gram equalities may have the same first-order variation and the rank can drop, causing this linear count to overestimate the family dimension. We therefore evaluate the rank at a generic point of the same exact family.

Table 1 in the main text and Fig. S4 summarize this classification. The displayed four-neighbor examples can be deformed continuously into one another so that, at every step, the two environments have equal two-decks and remain noncongruent. The six- and seven-neighbor examples instead belong to different families because they contain different numbers of neighbors. The cyclic examples are grouped by a common counting mechanism rather than by a continuous deformation. Among them, only the 9-of-18 ring I was found by both agents.

### A. Seven-parameter four-neighbor power family

In the complete-basis limit, the power spectrum contains the geometry of every unordered pair of neighbors. For a pair $\{\boldsymbol{r}_i, \boldsymbol{r}_j\}$, this geometry is completely specified up to $O(3)$ by the $2 \times 2$ Gram matrix

$$G(i, j) = \begin{pmatrix} |\boldsymbol{r}_i|^2 & \boldsymbol{r}_i \cdot \boldsymbol{r}_j \\ \boldsymbol{r}_i \cdot \boldsymbol{r}_j & |\boldsymbol{r}_j|^2 \end{pmatrix}.$$

Consequently, two environments have the same complete power spectrum if they contain the same multiset of these pairwise Gram matrices, including multiplicities. We do not need their single full Gram matrices to agree: that would generally make the entire environments congruent.

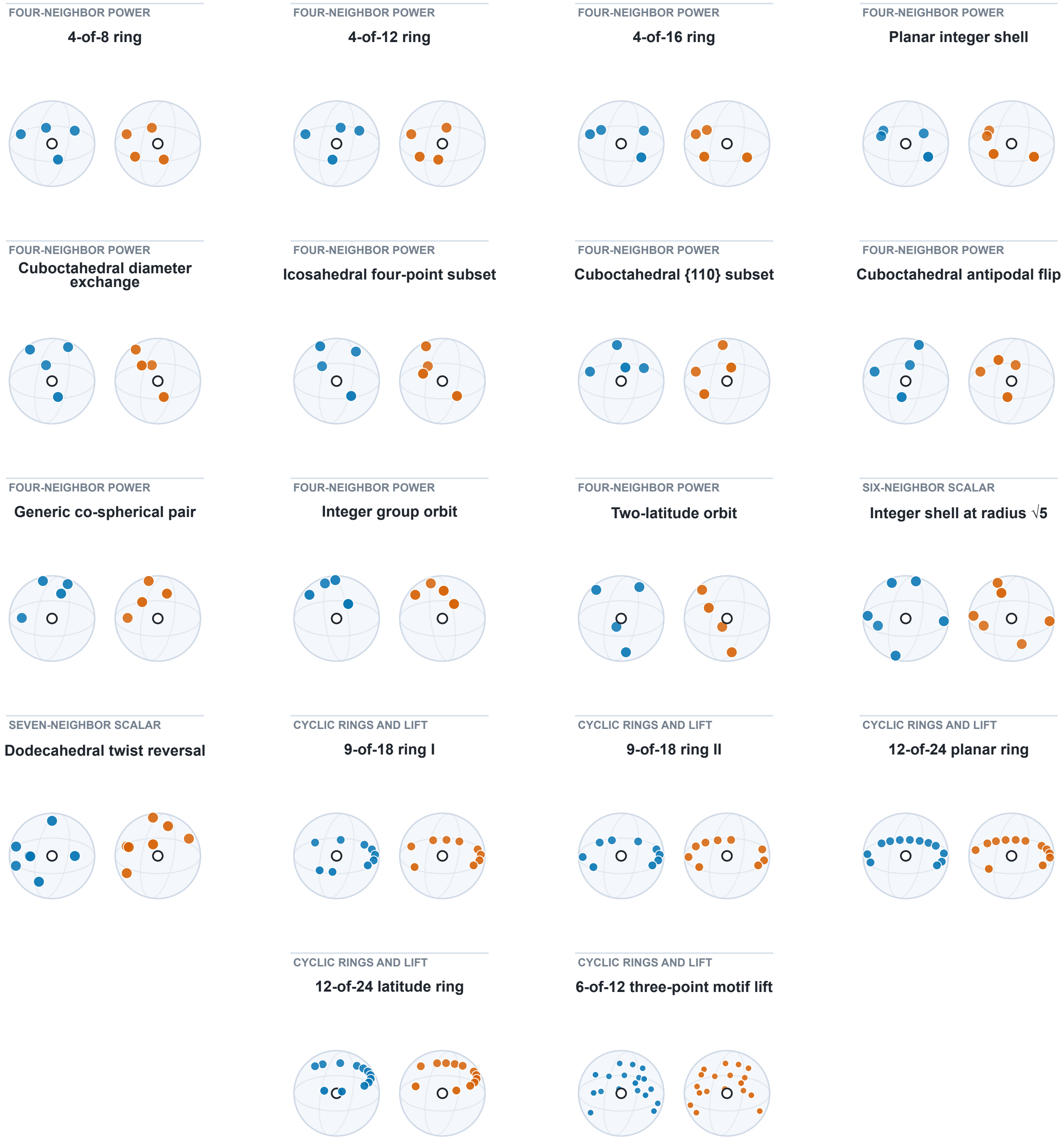


FIG. S4. Atlas of the representative classes returned by the repeated isolated searches. Blue and orange show the two environments in each pair; the open marker is the central atom. The family tags identify the four-neighbor power, six-neighbor scalar, seven-neighbor scalar, and cyclic rings and lift classes. Exact definitions are given in Table S2, Eqs. (S48) and (S52), and Table S3.

The parametrization below constructs this matching explicitly. After relabeling three common points and applying an orthogonal change of frame, every four-neighbor representative in Fig. S4 can be written as

$$\begin{aligned} \boldsymbol{r}_+ &= (0,a,b), & \boldsymbol{r}_- &= (0,-a,b), & & \\ \boldsymbol{r}_1 &= (c,0,d), & \boldsymbol{r}_2 &= (e,f,g), & \boldsymbol{r}_3 &= (-e,-f,g), \\ X(\boldsymbol{p}) &= \{\boldsymbol{r}_1,\boldsymbol{r}_2,\boldsymbol{r}_3,\boldsymbol{r}_+\}, & Y(\boldsymbol{p}) &= \{\boldsymbol{r}_1,\boldsymbol{r}_2,\boldsymbol{r}_3,\boldsymbol{r}_-\}, & & \end{aligned} \tag{S42}$$

with $\boldsymbol{p} = (a,b,c,d,e,f,g)$. The two environments share $\boldsymbol{r}_1, \boldsymbol{r}_2$, and $\boldsymbol{r}_3$, so every pair formed entirely from these three points has exactly the same Gram matrix in $X$ and $Y$. It remains only to match the three pairs containing $\boldsymbol{r}_+$ in $X$ with the three pairs containing $\boldsymbol{r}_-$ in $Y$.

First, $|\boldsymbol{r}_+| = |\boldsymbol{r}_-|$ and $|\boldsymbol{r}_2| = |\boldsymbol{r}_3|$. The pair containing $\boldsymbol{r}_1$ matches without exchanging any point because

$$\boldsymbol{r}_1 \cdot \boldsymbol{r}_+ = \boldsymbol{r}_1 \cdot \boldsymbol{r}_- = bd.$$

The other two pairs match after exchanging $\boldsymbol{r}_2$ and $\boldsymbol{r}_3$:

$$\begin{aligned} \boldsymbol{r}_2 \cdot \boldsymbol{r}_+ &= af + bg = \boldsymbol{r}_3 \cdot \boldsymbol{r}_-, \\ \boldsymbol{r}_3 \cdot \boldsymbol{r}_+ &= -af + bg = \boldsymbol{r}_2 \cdot \boldsymbol{r}_-. \end{aligned} \tag{S43}$$

Together with the equal norms, these identities show that the corresponding $2 \times 2$ Gram matrices are identical. Every pairwise card in $X$ therefore has a partner with the same multiplicity in $Y$. The complete two-decks, and hence the complete power spectra, are equal for every value of $\boldsymbol{p}$.

Table S2 gives the parameters of the eleven representatives in this common seven-parameter family. For every row, our systematic-classification validation script [2] reconstructs the displayed pair, verifies equality of the two-decks, and confirms that the two environments are not related by an $O(3)$ transformation.

Sharing the seven-parameter form in Eq. (S42) does not by itself establish this continuous connection, since a path could pass through an invalid or trivial configuration. We therefore vary the seven parameters linearly from each displayed example to the same reference point. The equal-deck identities above hold throughout, and our validation script uses exact polynomial calculations to check that no neighbor reaches the center or collides with another and that the two environments never become related by a rotation or reflection. For the generic numerical report, the script first aligns a fitted member of the family with the reported coordinates at their printed precision and then connects that fit to the nearby exact rational representative in Table S2.

We now apply the general dimension count above. Each four-neighbor environment has six two-neighbor cards. For every card matching established in the preceding equations, equality of the two $2 \times 2$ Gram matrices gives three scalar equations. If the matched cards are $\{\mathbf{u}, \mathbf{v}\} \subset X$ and $\{\mathbf{u}', \mathbf{v}'\} \subset Y$, these equations are

$$|\mathbf{u}|^2 = |\mathbf{u}'|^2, \qquad |\mathbf{v}|^2 = |\mathbf{v}'|^2, \qquad \mathbf{u} \cdot \mathbf{v} = \mathbf{u}' \cdot \mathbf{v}'.$$

The six card matches therefore produce 18 scalar Gram-entry equalities in the 24 Cartesian coordinates of the two environments. Some of these equations repeat the same norm constraint, so they are not all independent. At generic parameter values their Jacobian has rank 11, and the local solution set consequently has dimension $24 - 11 = 13$ before removing rotations. Six of these dimensions are the three rigid rotations of each environment, leaving

$$d = 24 - 11 - 6 = 7,$$

in agreement with the seven explicit parameters in Eq. (S42). This agreement shows both that the parametrization generates a seven-dimensional family and that there are no additional local shape parameters at a generic point. At some more symmetric representatives the Jacobian rank is smaller because several Gram-entry equations become indistinguishable to first order. Such a rank drop does not by itself imply an additional finite deformation, so the generic rank is the relevant quantity for the dimension of this family. The codimension is therefore $c = 9 - 7 = 2$.

### B. Six-neighbor scalar representative

In the complete-basis limit, the reflection-even scalar bispectrum records the multiset of three-neighbor geometries. Each such geometry is represented by a $3 \times 3$ Gram matrix, up to a simultaneous permutation of its rows and columns. Although the searches returned three six-neighbor reports, after removing their overall scale and orientation they all describe the same representative pair. This subsection studies that single pair: to determine the dimension of the equal-deck class passing through it, we allow its inner products to vary and show that the example has a four-parameter continuous extension.

| representative | $\boldsymbol{p} = (a, b, c, d, e, f, g)$ | Codex off | Codex on | Claude off | Claude on |
|---|---|---|---|---|---|
| 4-of-8 ring | $(1, 0, -1, 0, -1/\sqrt{2}, 1/\sqrt{2}, 0)$ | 4 | 0 | 6 | 3 |
| 4-of-12 ring | $(1, 0, -1, 0, -1/2, \sqrt{3}/2, 0)$ | 2 | 0 | 0 | 1 |
| 4-of-16 ring | $(17/10, 0, -17/10, 0, -17\sqrt{2+\sqrt{2}}/20, 17\sqrt{2-\sqrt{2}}/20, 0)$ | 0 | 0 | 1 | 0 |
| Planar integer shell | $(\sqrt{5}, 0, -\sqrt{5}, 0, -6/\sqrt{5}, 2/\sqrt{5}, 0)$ | 1 | 0 | 0 | 0 |
| Cuboctahedral diameter exchange | $(1, 1, -1, 1, -1, 1, 0)$ | 1 | 0 | 0 | 0 |
| Icosahedral four-point subset | $(\varphi, 1, -1, \varphi, -\varphi, 1, 0)$ | 0 | 1 | 0 | 0 |
| Cuboctahedral $\{110\}$ subset | $(\sqrt{2}, 0, -\sqrt{2}, 0, -1/\sqrt{2}, 1/\sqrt{2}, 1)$ | 0 | 1 | 0 | 1 |
| Cuboctahedral antipodal flip | $(\sqrt{2/3}, 2/\sqrt{3}, -\sqrt{2}, 0, -1/\sqrt{2}, \sqrt{3/2}, 0)$ | 0 | 1 | 0 | 0 |
| Generic co-spherical pair | $(30/29, 63/58, -189/130, -24/65, -19/30, 4/15, 4/3)$ | 0 | 0 | 0 | 1 |
| Integer group orbit | $(1, 2\sqrt{2}, -3/\sqrt{2}, 3/\sqrt{2}, -1/\sqrt{2}, 2, 3/\sqrt{2})$ | 0 | 0 | 1 | 0 |
| Two-latitude orbit | $(4/5, 3/5, -4/5, 3/5, -2\sqrt{2}/5, 2\sqrt{2}/5, -3/5)$ | 0 | 0 | 0 | 1 |

TABLE S2. Representatives of the four-neighbor power family, in the gauge of Eq. (S42). The four rightmost columns count distinct pairs returned by the Codex and Claude searches under the offline (off) and online (on) conditions. Repetitions across separate runs are retained, but duplicate outputs within one run are counted once. Here $\varphi = (1+\sqrt{5})/2$ is the golden ratio. The generic row gives a nearby exact rational representative of the component associated with the numerical Claude-online report through the fit described above: its four points lie on the radius-3/2 sphere.

Begin with three vectors $\mathbf{u}, \mathbf{v}, \mathbf{w}$ satisfying

$$\begin{gathered} \mathbf{u}\cdot\mathbf{u} = h, \qquad \mathbf{v}\cdot\mathbf{v} = \mathbf{w}\cdot\mathbf{w} = e, \\ \mathbf{u}\cdot\mathbf{v} = a, \qquad \mathbf{u}\cdot\mathbf{w} = b, \qquad \mathbf{v}\cdot\mathbf{w} = 0, \end{gathered} \tag{S44}$$

where $h > 0$, $e > 0$, and $\Delta = he - a^2 - b^2 > 0$. These conditions make the Gram matrix of $\mathbf{u}, \mathbf{v}, \mathbf{w}$ positive definite, so the three vectors are linearly independent. A vector in their span is therefore uniquely determined by its inner products with $\mathbf{u}, \mathbf{v}, \mathbf{w}$.

Set $c = \Delta/e$ and define $\mathbf{s}$ and $\mathbf{t}$ by

$$\begin{aligned} \mathbf{u}\cdot\mathbf{s} &= c, \;\; \mathbf{v}\cdot\mathbf{s} = b, \quad \mathbf{w}\cdot\mathbf{s} = -a, \\ \mathbf{u}\cdot\mathbf{t} &= c, \;\; \mathbf{v}\cdot\mathbf{t} = -b, \;\; \mathbf{w}\cdot\mathbf{t} = a. \end{aligned} \tag{S45}$$

The particular value of $c$ is chosen so that $|\mathbf{s}|^2 = |\mathbf{t}|^2 = h$. We can therefore form

$$X = \{\mathbf{u}, \mathbf{v}, \mathbf{w}, \mathbf{s}, -\mathbf{s}, -\mathbf{u}\}, \qquad Y = \{\mathbf{u}, \mathbf{v}, \mathbf{w}, \mathbf{t}, -\mathbf{t}, -\mathbf{u}\} \tag{S46}$$

and compare their three-neighbor cards.

Every three-neighbor card contains either neither member of the added antipodal pair, exactly one member, or both. The four points $\mathbf{u}, \mathbf{v}, \mathbf{w}, -\mathbf{u}$ are common to both environments, so cards of the first type are unchanged. To compare the other two types, all relevant inner products are collected in the following table:

$$\begin{array}{c|cccc} & \mathbf{u} & \mathbf{v} & \mathbf{w} & -\mathbf{u} \\ \hline \mathbf{s} & c & b & -a & -c \\ -\mathbf{s} & -c & -b & a & c \\ \mathbf{t} & c & -b & a & -c \\ -\mathbf{t} & -c & b & -a & c \end{array} \tag{S47}$$

Together with $|\mathbf{u}| = |\mathbf{s}| = |\mathbf{t}|$ and $|\mathbf{v}| = |\mathbf{w}|$, this table pairs every card containing exactly one added point. For example, $\{\mathbf{v}, \mathbf{w}, \mathbf{s}\}$ in $X$ has exactly the same Gram matrix as $\{\mathbf{v}, \mathbf{w}, -\mathbf{t}\}$ in $Y$, while $\{\mathbf{u}, \mathbf{v}, \mathbf{s}\}$ matches $\{\mathbf{t}, \mathbf{w}, \mathbf{u}\}$ after the three atoms are relabeled. Cards containing both $\mathbf{s}$ and $-\mathbf{s}$ match those containing both $\mathbf{t}$ and $-\mathbf{t}$, because the antipodal pairs have the same radius and the same two inner products, with opposite signs, with every common point. This covers all three cases.

It follows that $X$ and $Y$ have the same multiset of $3 \times 3$ Gram matrices, including the same multiplicities, for every admissible $(h, e, a, b)$. Our systematic-classification validation script [2] verifies this directly by enumerating all $\binom{6}{3} = 20$ cards. The integer representative shown in the atlas is obtained from

$$(h, e, a, b) = (5, 5, 4, 2). \tag{S48}$$

At this point $c = 1$, and one possible realization is

$$\mathbf{u} = (-2, -1, 0), \quad \mathbf{v} = (-2, 0, -1), \quad \mathbf{w} = (-1, 0, 2), \quad \mathbf{s} = (0, -1, -2), \quad \mathbf{t} = (0, -1, 2).$$

Varying $(h, e, a, b)$ within the inequalities above gives four independent shape parameters. The Jacobian of the equalities that match the 20 three-neighbor Gram matrices has generic rank 26 in the 36 coordinates of the two environments. After removing the three rotational degrees of freedom of each environment, this gives $d = 36 - 26 - 6 = 4$, consistently with the explicit construction, and hence codimension $c = 15 - 4 = 11$.

This equality concerns the reflection-even scalar bispectrum. A Gram matrix determines only the magnitude of the scalar triple product, not its sign. The corresponding signed-volume channels distinguish $X$ from $Y$.

### C. Seven-neighbor scalar representative

The two retained local dodecahedral reports reduce, after removing their overall scale and orientation, to the same twist-reversal pair. To determine the dimension of the equal-deck class passing through it, we place the example in a six-parameter continuous extension, with one point on an axis and the other six on two equal-radius rings. In a rotation gauge,

$$\begin{aligned}
X_1 &= (0, 0, L), \\
X_0 &= (r\cos\psi, r\sin\psi, h), \\
X_2 &= (r\cos(\psi + u), r\sin(\psi + u), -h), \\
X_5 &= (r\cos(\psi + v), r\sin(\psi + v), -h), \\
X_6 &= (r\cos(\psi + w), r\sin(\psi + w), h), \\
X_4 &= (r\cos(\psi + v - w), r\sin(\psi + v - w), -h), \\
X_3 &= (r\cos(\psi + u + w - v), r\sin(\psi + u + w - v), h).
\end{aligned} \tag{S49}$$

The angle $\psi$ is a global rotation and the six shape parameters are $\boldsymbol{q} = (L, h, r, u, v, w)$. To construct the partner, reverse the axial vector $X_1$, and then apply the same orthogonal transformation

$$Q(x, y, z) = (z, x, -y) \tag{S50}$$

to all seven points. As an unordered environment, the result is therefore

$$Y = Q\{X_0, -X_1, X_2, X_3, X_4, X_5, X_6\}. \tag{S51}$$

Since $Q$ preserves inner products, applying $Q^{-1}$ to $Y$ does not change its Gram deck. We may therefore compare $X$ directly with $Q^{-1}Y = \{X_0, -X_1, X_2, X_3, X_4, X_5, X_6\}$. These environments differ only in the sign of the axial point $X_1$, so every card not containing $X_1$ is unchanged. Consider now a three-card containing $X_1$ and two ring points. If one point lies on each ring, reversing $X_1$ exchanges its two axial inner products, $+hL$ and $-hL$. The card is therefore preserved by exchanging the labels of the two ring points.

It remains to consider two points on the same ring. After removing the common angle $\psi$, the upper-ring points $X_0, X_6, X_3$ have angles $0, w, u + w - v$, respectively, while the lower-ring points $X_2, X_5, X_4$ have angles $u, v, v - w$. Match the three upper-ring pairs with lower-ring pairs as follows:

$$\begin{aligned}
\{X_0, X_6\} &\longleftrightarrow \{X_4, X_5\}, \quad \Delta\theta = w, \\
\{X_0, X_3\} &\longleftrightarrow \{X_2, X_4\}, \quad \Delta\theta = u + w - v, \\
\{X_6, X_3\} &\longleftrightarrow \{X_2, X_5\}, \quad \Delta\theta = u - v.
\end{aligned}$$

In each row, the two pairs have the same angular difference up to sign and hence the same mutual inner product, $r^2\cos(\Delta\theta) + h^2$. Moreover, $X_1$ has inner product $+hL$ with every upper-ring point, while $-X_1$ has the same inner product with every lower-ring point. Thus, every card containing $X_1$ and two upper-ring points matches the card containing $-X_1$ and the corresponding lower-ring pair. For a card containing $X_1$ and two lower-ring points, use the same correspondence in the opposite direction, matching it with the card containing $-X_1$ and the corresponding upper-ring pair. The two axial inner products are then $-hL$, and the mutual inner product of the ring points is unchanged. These three cases establish equality of the complete scalar three-decks for every $\boldsymbol{q}$. Varying $\boldsymbol{q}$ gives an exact equal-pair path. The dodecahedral twist-reversal point shown in the atlas is

$$\boldsymbol{q} = \left(\sqrt{3}, \frac{1}{\sqrt{3}}, \sqrt{\frac{8}{3}}, \frac{\pi}{3}, -\arctan\frac{\sqrt{3}}{4 + \sqrt{5}}, \arctan\frac{\sqrt{3}}{4 - \sqrt{5}}\right). \tag{S52}$$

At a generic exact point of this family, the matched Jacobian has rank 30 in 42 pair coordinates. After removing six independent rotations, the family dimension is $42 - 30 - 6 = 6$ and its codimension is $18 - 6 = 12$.

The dodecahedral view hides a simpler organization that is already present in Eq. (S49) and is analogous to the one presented in Ref. [12]. Appropriately rotated, the points $X_0, X_6, X_3$ form a triangle in the plane $z = h$, while $X_2, X_5, X_4$ form a second triangle in the parallel plane $z = -h$. All six lie on rings of radius $r$ about the same axis, and $X_1$ lies on that axis. Figure S5 shows this directly.

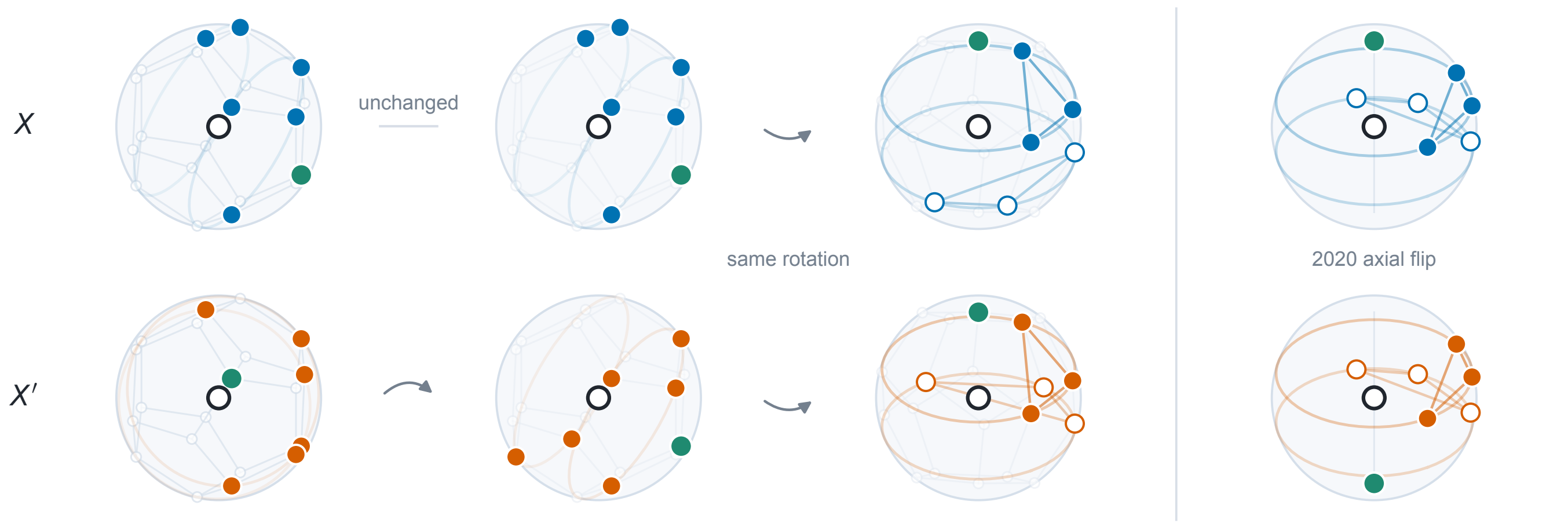


FIG. S5. Reorienting the dodecahedral pair and comparing it with the 2020 construction. In the first column, the seven selected neighbors are shown on the faint full dodecahedral shell; the green point identifies the neighbor that becomes axial. In the second column, only $X'$ is rotated so that its ring axis and green point align with those of $X$. The same additional rotation is then applied to both environments, giving the third column: the green point is at the north pole in both rows, while the two triangles on the parallel rings have opposite relative twists. These transformations are all proper rotations. The final column shows the bispectrum counterexample of Ref. [12], in the same orientation and with the spherical rendering used in Fig. 1b of the main text. Here the two triangles remain fixed and the green axial point moves from the north to the south pole. Filled and open markers distinguish the two ring planes.

Figure S5 suggests a more transparent description of the same family. Let $\boldsymbol{\beta} = (\beta_1, \beta_2, \beta_3)$ give the angles of the upper triangle, and let $\alpha$ be the relative rotation of the lower triangle. We denote the two possible twist directions by

$$\begin{aligned} \mathbf{a}_j &= (R\cos\beta_j, R\sin\beta_j, h), \\ \mathbf{b}_j^{(\pm)} &= \big(R\cos(\beta_j \pm \alpha), R\sin(\beta_j \pm \alpha), -h\big). \end{aligned} \tag{S53}$$

We write $\mathbf{T}_\pm = \{\mathbf{a}_1, \mathbf{a}_2, \mathbf{a}_3, \mathbf{b}_1^{(\pm)}, \mathbf{b}_2^{(\pm)}, \mathbf{b}_3^{(\pm)}\}$ for the corresponding six triangle points. This introduces no new construction. After removing the global angle $\psi$, direct comparison with Eq. (S49) gives $R = r$, $\boldsymbol{\beta} = (0, w, u + w - v)$, and $\alpha = v - w$; the heights $h$ and $L$ are unchanged. Substituting Eq. (S52) therefore gives exactly the dodecahedral pair shown in Fig. S5.

Because the scalar three-deck equality holds for every $\boldsymbol{q}$, this dodecahedral point can be varied continuously within the same six-parameter degenerate family. One convenient endpoint is the rational geometry

$$(R, h, L) = \left(1, \frac{2}{5}, \frac{7}{5}\right), \qquad \boldsymbol{\beta} = \left(0, \frac{\pi}{2}, \arctan\frac{4}{3}\right), \qquad \alpha = \arctan\frac{12}{5}. \tag{S54}$$

At this endpoint, the family reverses the relative twist while keeping the axial point fixed, whereas the 2020 counterexample keeps the triangles fixed and flips the axial point:

$$\begin{aligned} \text{twist reversal:} \quad & \mathbf{T}_+ \cup \{(0,0,L)\} \longleftrightarrow \mathbf{T}_- \cup \{(0,0,L)\}, \\ \text{2020 axial flip:} \quad & \mathbf{T}_+ \cup \{(0,0,L)\} \longleftrightarrow \mathbf{T}_+ \cup \{(0,0,-L)\}. \end{aligned} \tag{S55}$$

The two lines share their first environment. Their second environments are mirror images. To state the relation explicitly, let $R_z(\theta)$ denote the proper rotation through angle $\theta$ about the $z$ axis, and let $-I$ denote inversion through the origin. We define

$$P_\alpha = R_z(\alpha - \pi)(-I).$$

| representative | $S_A$ | $S_B$ | Codex | | Claude | |
|---|---|---|---|---|---|---|
| | | | off | on | off | on |
| 9-of-18 ring I | $\{0,1,2,3,4,6,8,14,15\}$ | $\{0,1,2,3,5,6,7,9,13\}$ | 4 | 0 | 1 | 1 |
| 9-of-18 ring II | $\{0,1,2,3,5,7,8,11,13\}$ | $\{0,1,3,6,7,8,9,11,13\}$ | 1 | 0 | 0 | 0 |
| 12-of-24 planar ring | $\{0,1,3,5,6,7,8,9,10,11,14,16\}$ | $\{0,2,3,4,5,7,8,9,10,11,13,18\}$ | 1 | 0 | 0 | 0 |
| 12-of-24 latitude ring | $\{0,1,2,3,4,5,6,8,10,11,19,21\}$ | $\{0,1,2,3,4,6,7,8,9,11,17,22\}$ | 0 | 0 | 1 | 0 |
| 6-of-12 three-point motif lift | $\{0,1,3,4,5,8\}$ | $\{0,3,4,5,7,8\} = 8 - S_A$ | 0 | 1 | 0 | 0 |

TABLE S3. Cyclic selectors for the five representatives in the final row of Fig. S4. The first three rows are planar rings, the fourth is the latitude ring, and the last is the three-point motif lift. The four rightmost columns give the Codex and Claude counts in the offline (off) and online (on) conditions.

Thus, $P_\alpha$ first inverts every point and then rotates the result through $\alpha - \pi$ about the $z$ axis. It is an improper orthogonal transformation, with $\det P_\alpha = -1$, and maps the twist-reversed structure onto the axial-flipped one:

$$P_\alpha(\mathbf{T}_- \cup \{(0,0,L)\}) = \mathbf{T}_+ \cup \{(0,0,-L)\}. \tag{S56}$$

Thus, the dodecahedral report and the 2020 example belong to the same scalar/$O(3)$ family. A generic twist-reversal pair is nevertheless distinguished by the parity-odd components of the $SO(3)$ bispectrum, because the correspondence above is improper and reverses signed volumes. At the axial-flip configuration the signed-volume deck also agrees, so this pair remains bispectrum-degenerate even when all orientation-sensitive channels are included.

### D. Cyclic rings and lift

The final search class contains five distinct pairs: three planar rings, one latitude ring, and one three-dimensional motif lift. They share the cyclic counting mechanism of Sec. S1, but do not form one continuous family: the ring size $N$ and selectors $S_A, S_B \subset \mathbb{Z}_N$ are discrete choices rather than shape parameters. The main-text table therefore groups these examples by mechanism and leaves their family dimension $d$ and codimension $c$ undefined. Table S3 lists the selectors and records where each pair was found.

For the four single-ring pairs, we write

$$\mathcal{R}_N(S; R, z) = \left\{ \left( R\cos\frac{2\pi s}{N}, R\sin\frac{2\pi s}{N}, z \right) : s \in S \right\}. \tag{S57}$$

**Planar rings.** The first three rows use $R = 1$ and $z = 0$. A cyclic shift of $S$ rotates the environment about the $z$ axis, while reversing its cyclic order is a rotation through $\pi$ about a suitable axis in the ring plane. The reflection-symmetrized cyclic counts therefore determine the complete geometric decks. For all three pairs they agree through order three and differ at order four.

**Latitude ring.** The fourth row keeps the same cyclic indexing but places the ring above the central atom, with

$$(R, z) = \frac{3}{2}\left(\cos\frac{3\pi}{20}, \sin\frac{3\pi}{20}\right).$$

All twelve neighbors remain at distance 3/2 from the center. The nonzero height changes the role of cyclic reversal: the planar rotation described above would move this ring from $z$ to $-z$, whereas reversing the order at fixed $z$ is an improper reflection. The cyclic orientation of a three-atom card can now be detected by the sign of its scalar triple product, so reflection-symmetrized counts are not sufficient. Instead, the two selectors satisfy

$$d_r^{S_A}(\boldsymbol{s}) = d_r^{S_B}(\boldsymbol{s}) \quad \text{for every } \boldsymbol{s} \text{ and } r = 1, 2, 3.$$

This directed equality fixes both the geometry and orientation of every card through order three while the four-decks differ.

**Three-dimensional motif lift.** The final row converts each selected cyclic index into a complete rotated copy of the fixed three-atom motif

$$\mathbf{m}_0 = (1, 0, 1/5), \quad \mathbf{m}_1 = (6/5, 1/5, -2/5), \quad \mathbf{m}_2 = (4/5, -3/10, 7/10), \qquad \mathcal{M} = \{\mathbf{m}_0, \mathbf{m}_1, \mathbf{m}_2\}. \tag{S58}$$

Let $R_j$ be the rotation about the $z$ axis through $2\pi j/12$. The twelve sets $R_j\mathcal{M}$, with $j = 0,\ldots,11$, are the twelve possible rotated copies of the motif. The cyclic selector $S$ chooses six of these copies, and the corresponding environment is

$$X_S = \bigcup_{j\in S} R_j\mathcal{M} = \{R_j\mathbf{m}_t : j \in S,\ t = 0, 1, 2\}. \tag{S59}$$

The two environments in the counterexample are $X_{S_A}$ and $X_{S_B}$. Each selected copy contributes three atoms, so each environment contains $6 \times 3 = 18$ atoms.

Viewed one atom at a time, the rotated images $R_j\mathbf{m}_t$ of each fixed $\mathbf{m}_t$ lie on one coaxial ring. The complete environment therefore lies on three rings. Crucially, these rings are not populated independently: selecting an index $j$ includes the whole rotated motif $R_j\mathcal{M}$, with one atom on each ring. The same selector $S$ thus controls every within-ring and cross-ring pattern.

To see this explicitly, write $\mathbf{r}_{j,t} = R_j\mathbf{m}_t$. A proper rotation by $R_{-j_1}$ sends any three-atom card to

$$\{\mathbf{r}_{j_1,t_1}, \mathbf{r}_{j_2,t_2}, \mathbf{r}_{j_3,t_3}\} \longmapsto \{\mathbf{m}_{t_1}, R_{j_2-j_1}\mathbf{m}_{t_2}, R_{j_3-j_1}\mathbf{m}_{t_3}\}.$$

Its geometry therefore depends only on the three motif labels $(t_1, t_2, t_3)$ and the directed cyclic differences $(j_2-j_1, j_3-j_1)$. Cards drawing atoms from three, two, or one distinct motif copies are counted by $d_3$, $d_2$, or $d_1$, respectively. These directed counts agree for $S_A$ and $S_B$, so every signed $SO(3)$ three-card has the same multiplicity in the two lifted environments. Their four-decks differ.

Our systematic-classification validation script [2] independently verifies the deck equalities and first differing orders of the displayed representatives. It reconstructs the eight exact power-source geometries and the dodecahedral source pair and aligns the generic fit with the reported coordinates within their printed precision. It also checks the continuous paths between the power-family representatives, confirms the rational two-triangle equivalence, and evaluates the three Jacobian ranks quoted above.

### E. Prompts used in the isolated searches

Each agent was launched in an isolated working directory containing the same collection of pre-2020 papers and one of the two instruction files reproduced below. The offline condition explicitly prohibited internet access, whereas the online condition encouraged a broader interdisciplinary literature search while retaining the same publication-date restriction. Each condition was repeated eight times with each agent. The text is reproduced verbatim from the corresponding `AGENTS.md` files.

**Offline condition**

```
Completeness of representations for atomic-scale structure
==========================================================

General instructions
--------------------

Base your reasoning on the papers in ‘literature/‘. Do not for any
reason rely explicitly on more recent work, from 2020 included
onwards. Do not look for literature, do not access internet nor
perform online search. Do not use any memories from our previous
interactions, nor from data outside of the present project folder.

The problem
-----------

The question is whether the low-order invariants, e.g. the SOAP
power spectrum and the bispectrum, cf. Bartok 2013, are "complete"
descriptors, i.e. if they are guaranteed to differentiate between
any atomic geometry that are not related by symmetry. In other
words, if there is a bijective map between the local atomic
point-cloud configuration and the descriptors. That this is not
obvious is clear by the existence of "homometric" structures that
have the same set of interatomic distances (equivalent to the
radial part of the density expansion coefficients) but are not
superimposable.
I am especially interested in structures that cannot be
distinguished regardless of the resolution (truncation of the
density expansion basis) and the cutoff (truncation of the
environment). Is the power spectrum (2-neighbors) enough? Is the
bispectrum? Is the trispectrum (4-neighbors?). Find the
highest-possible order of degeneracy for a structure, isolated or
periodic, relative to one or all of its centers.

Required output
---------------

Generate a pair of structures, saving them as ‘<prefix>-a.xyz‘ and
‘<prefix>-b.xyz‘ for each distinct construction that you identify,
together with a ‘<prefix>.md‘ file describing the logic of the
construction and providing the verified references that you have
used, if any. If the two structures are supposed to be degenerate
only relative to one "special center" make that the first atom and
call it C, and call H all the other atoms that should be only
neighbors. If a structure is degenerate relative to all atoms,
then make all atoms C. Work exclusively in the present folder and
keep a dump of the conversation in there as well.
```

**Online condition**

```
Completeness of representations for atomic-scale structure
==========================================================

General instructions
--------------------

Base your reasoning on the papers in ‘literature/‘. Do not for any reason rely
explicitly on more recent work, from 2020 included onwards. Look for interdisciplinary
references, also outside the specific application domain, and on your own deductions
based on the available literature. Do not use data outside of the present project
folder.

Context
-------

Machine-learning interatomic potentials have been built, at least starting from the
early 2000s, based on the construction of atom-centered descriptors of the geometry
and composition of molecules and materials. In order to ensure that the model is
consistent with Euclidean symmetries (translations, continuous rotations) and with
permutations of atomic indices, one typically starts by building an expansion of the
neighbor density in spherical harmonics and radial functions, and then combines them
to obtain rotationally-invariant coefficients.
The logic is discussed extensively in the reference papers in ‘literature/‘.


The problem
-----------

The question is whether the low-order invariants, e.g. the SOAP power spectrum and the
bispectrum, cf. Bartok 2013, are "complete" descriptors, i.e. if they are guaranteed
to differentiate between any atomic geometry that are not related by symmetry. In
other words, if there is a bijective map between the local atomic point-cloud
configuration and the descriptors. That this is not obvious is clear by the existence
of "homometric" structures that have the same set of interatomic distances (equivalent
to the radial part of the density expansion coefficients) but are not superimposable.
I am especially interested in structures that cannot be distinguished regardless of
the resolution (truncation of the density expansion basis) and the cutoff (truncation
of the environment). Is the power spectrum (2-correlations) enough? Is the bispectrum
(3-correlations)? Is the trispectrum (4-correlations?) Is the tetraspectrum
(5-correlations?). Find the highest-possible order of degeneracy for a structure,
isolated or periodic, relative to one or all of its centers.
While planar examples are considered succesful findings already, I am interest in
counterexamples with a full 3D point cloud which do not rely on a rank-2 construction.
This does not include the trivial case in which we add two symmetric points at the
north and south pole of the point cloud to lift the proof from planar to 3D. Instead,
we should look for a genuine different geometry that is not fundamentally based on a
planar polygon.

Required output
---------------

Generate a pair of structures, saving them as ‘<prefix>-a.xyz‘ and ‘<prefix>-b.xyz‘
for each distinct construction that you identify, together with a ‘<prefix>.md‘ file
describing the logic of the construction and providing the verified references that
you have used, if any. If the two structures are supposed to be degenerate only
relative to one "special center" make that the first atom and call it C, and call H
all the other atoms that should be only neighbors. If a structure is degenerate
relative to all atoms, then make all atoms C. Work exclusively in the present folder
and keep a dump of the conversation in there as well.
```


\AddToHookNext{shipout/before}{\DiscardShipoutBox}

\end{document}